\documentclass[twocolumn,numberedappendix, twocolappendix, apj]{openjournal}

\usepackage{amsfonts,textcomp}
\usepackage[T1]{fontenc}
\usepackage{xcolor}
\usepackage[normalem]{ulem}

\usepackage{graphicx}	
\usepackage{amsmath}	
\usepackage{amssymb}	
\usepackage{bm}         
\usepackage{float}

\begin{document}

\author[0000-0002-7537-6921]{Aoife Boyle$^{1,\star}$}
\author[0000-0003-1060-3959]{Alexandre Barthelemy$^{2}$}
\author{Sandrine Codis$^{1}$}
\author[0000-0001-7831-1579]{Cora Uhlemann$^{3}$}
\author{Oliver Friedrich$^{2}$}
\thanks{$^\star$\href{mailto:aoife.boyle@cea.fr}{aoife.boyle@cea.fr}}

\affiliation{$^{1}$ AIM, CEA, CNRS, Université Paris-Saclay, Université Paris Diderot, Sorbonne Paris Cité, 91191 Gif-sur-Yvette, France}
\affiliation{$^{2}$ Universitäts-Sternwarte, Fakultät für Physik, Ludwig-Maximilians Universität München, Scheinerstr. 1, 81679 München, Germany}
\affiliation{$^{3}$ School of Mathematics, Statistics and Physics, Newcastle University, Herschel Building, NE1 7RU Newcastle-upon-Tyne, U.K.}

\title{The cumulant generating function as a novel observable\\ to cumulate weak lensing information}


\begin{abstract}
Key non-Gaussian properties of cosmological fields can be captured by their one-point statistics, providing a complement to two-point statistical measurements from power spectra or correlation functions. Large deviation theory can robustly predict the one-point statistics of cosmological density fields on mildly non-linear scales from first principles. It provides a direct prediction for the cumulant generating function (CGF) of such fields, from which a prediction for the more commonly used probability density function (PDF) is extracted through an inverse Laplace transform. For joint one-point statistics of multiple fields, the inverse Laplace transform rapidly becomes more cumbersome and computationally expensive. In this work, we demonstrate for the first time that the weak lensing CGF itself can be used as an observable that captures an equal amount of cosmological information to the PDF. While we use the weak-lensing convergence field as a simplistic and instructive example, this work is intended as a first step towards a cosmological analysis based on large deviation theory in the context of a nulling framework, which excludes contributions from small scales to facilitate highly accurate theoretical predictions. In this context, the method should be generally applicable for a multi-scale tomographic analysis of weak lensing and galaxy clustering. 
\end{abstract}

\maketitle




\section{Introduction}

Squeezing the maximum amount of cosmological information out of galaxy surveys with summary statistics requires statistics that go beyond the standard two-point correlation functions or power spectra, which only contain complete statistical information for Gaussian random fields. Most recently this has been exemplified by the exploitation of the Dark Energy Survey \citep[see for example][]{2022MNRAS.511.2075Z,2022PhRvD.105j3537S,gattiDES,GruenDES17}, and numerous papers \citep[see for example][and references therein]{2021JCAP...01..028Z,2022arXiv220601450Z} that have underlined the importance of non-Gaussian statistics for upcoming surveys like Euclid \citep{Euclid} or the Vera Rubin Observatory (LSST) \citep{LSST}. One-point statistics are of particular interest because of their ease of measurement and theoretical accessibility. In particular, the moments of the density field have received a lot of attention from both a theoretical 
\citep[e.g][]{1984ApJ...279..499F,1995MNRAS.274..730L,1995A&A...296..575B} and observational \citep[see][and references therein]{2013MNRAS.435....2W,gattiDES} point of view. 
Going one step further, several works have shown that the traditional one-point probability density function (PDF) is a particularly relevant additional non-Gaussian probe, which contains information on the full set of higher-order moments \citep{2016MNRAS.460.1549C,2020MNRAS.495.4006U,2021arXiv210702300F,Boyle21}. One other notable advantage of the PDF is that cutting out the extreme tails allows removal of the impact of rare events, which are 
difficult to account for properly in a cosmological data analysis.
In this paper, we advocate for the cumulant generating function (CGF) as a new observable analogous to the PDF, with desirable properties for straightforward extensions to multiple fields.

The cumulant generating function is a widely used mathematical object, closely related to the moment generating function (MGF), whose successive derivatives in zero are the cumulants (resp. moments) of the random variable under consideration. The CGF is actually defined as the natural logarithm of the MGF. From a mathematical viewpoint, cumulants and the CGF are preferred over moments and the MGF for problems that consider sums of independent (multi-variate) random variables, because the CGF of the sum in such cases is simply the sum of the CGFs. More concretely, cumulants represent the connected part of the corresponding moments. This means that the $n^{\rm th}$ cumulant gives extra information not contained in any previous cumulant, while additional moments can exist without adding any extra information. A typical example of the latter case is found in Wick's theorem for Gaussian fields, which quantitatively expresses all the higher order moments as a function of the first two (the mean and variance). The same Gaussian field, however, only admits two non-zero cumulants. This gives an additional advantage to considering cumulants in cosmology. In this context, the fields we consider are typically mildly non-Gaussian, so the higher-order cumulants are small and become smaller as their degree increases \citep[a natural consequence of the process of gravitational instability on cosmological scales, as shown extensively by, for example][]{Peebles,1995MNRAS.274.1049B,BernardeauReview}. 

From a physical viewpoint, CGFs are ubiquitous, especially in the realm of statistical physics where typically any extensive quantity can be related to the cumulants of a random variable (often the energy of the system). It can be noted, for example, that Helmholtz free energy is exactly the CGF of the energy and as such all derivatives of the free energy, such as the internal energy, entropy, and specific heat capacity, can be readily expressed in terms of cumulants. It is thus no surprise that the CGF is at the heart of the mathematical theory of large deviations (see for example \cite{touchette} for an introduction and the link with statistical mechanics and more historical references, or \cite{Ellis_book} for an earlier but more complete reference), 
which can be seen from a practical point of view as the mathematical formalism that underpins statistical mechanics. 

In cosmology, moment or cumulant generating functions were introduced in theoretical papers in the late 1970s as a way to probe some part of high-order correlation functions of clustering fields, which are hard to measure in practical experiments \citep[see for example][on using the probability of voids as a generating function for all other probabilities]{White1979}. Many attempts were made to compute these clustering functions and compare the results with numerical simulations in the following decade \citep[for a detailed discussion of the advances at the time, see the introduction of][]{Balian_1989}. We highlight in particular \cite{Balian_1989}, who showed how the existence of a critical point in the matter density CGF induces an exponential cut-off in the corresponding PDF, and \citep{Bernardeau_1992}, who made the link between the tree-order cumulant hierarchy in Eulerian perturbation theory and the dynamics of spherical collapse.
In recent years, and as a generalisation of the previous results using Eulerian perturbation theory, it has been shown that large-deviation theory (hereafter LDT) can be used reliably to access the one-point statistics of the cosmic density field on mildly nonlinear scales\footnote{Beyond the mildly nonlinear regime, phenomenological approaches based on the halo model \citep{Thiele20},  hierarchical or lognormal models \citep{Valageas_2004,Barber_2004,Repp_2017,Bernardeau_2022,Uhlemann2022} are needed.}\citep{seminalLDT}. Through projection, it can be connected to observables such as cosmic shear fields. A natural approach is to rely directly on the CGF rather than the PDF, because the theoretical prediction for the latter must be obtained from the CGF through a (not always tractable) complicated inverse Laplace transform. Moreover, cosmological analyses of real data, relying for example on Monte-Carlo Markov chains, require a huge number of evaluations of a theoretical model for different cosmologies, reinforcing the importance of efficiency.

In this work, we focus on exploring the cosmological information content of the CGF of the weak lensing convergence, but the results should be generally applicable to other cosmological fields. For weak lensing and photometric clustering observables, the CGF is particularly advantageous. The widely applied Limber approximation assumes that the correlations of the underlying matter density field along the line of sight are negligible compared to those in the transverse directions, which means that individual redshift slices can be treated as statistically independent \citep{BernardeauValageas,Barthelemy20a}. This makes the CGF a much more practical statistical tool than the PDF, because the CGFs in the individual redshift slices can simply be summed up. This is to be contrasted with the PDF, as \cite{Barthelemy22} showed that the computational cost of calculating joint PDFs between redshift bins of source galaxies from large deviation theory rapidly increases with the number of redshift bins, to the point of becoming intractable for the exploitation of upcoming surveys like Euclid \citep{Euclid}. 

The main goal of this paper is to demonstrate how the CGF can be exploited to capture the same cosmological information as the corresponding PDF (in the same way that the two-point correlation function and the power spectrum should lead to identical cosmological constraints), and to perform a simple cosmological Fisher forecast for the weak lensing convergence field. The goal is to provide a preliminary proof of concept for eventual joint CGF analyses of lensing and clustering fields. In this example, we use the large deviations formalism to predict the theoretical convergence CGF and the corresponding PDF. In recent work, \cite{characteristic_function} exploited an estimator based on the characteristic function to demonstrate its potential for extracting cross correlations between fields. The characteristic function is the Fourier transform of the PDF,  that is, the MGF with a purely imaginary parameter, so the two approaches are not dissimilar. 

Although, similarly to the CGF and with similar references, the history of the PDF in cosmology is quite old, recent motivation for the application of PDFs can be found in, for example, \cite{Boyle21}, in which the authors made use of a Fisher analysis approach to perform a forecast for the LDT-based theoretical model of the convergence PDF. It was found that the convergence PDF yields significantly tighter constraints on cosmological parameters, including the equation of state of dark energy and the neutrino mass, than the two-point correlation function. \cite{Patton17} used numerical simulations to demonstrate that the weak lensing convergence PDF provides information complementary to that of the cosmic shear two-point correlation function, also in the presence of weak lensing systematics. This complementarity between two-point and non-Gaussian statistics is even more relevant in the presence of shape noise. \cite{Liu19WLPDF} also used numerical simulations to demonstrate the power of the tomographic weak lensing PDF in helping to constrain the total neutrino mass. 

From the perspective of theoretical modelling, there exist numerous works based on numerical simulations or, more recently, the halo model as in \cite{Thiele20}. Such approaches have the advantage of being appropriate for modelling smaller scales but require phenomenological input. We will focus instead here on "from first principles" approaches that are applicable to somewhat larger scales. 
As early as \cite{BernardeauValageas} and its follow-up papers, hierarchical models for the statistics of the matter field were used to construct the PDF of the aperture mass. These works were later re-interpreted in terms of a large deviation principle, paving the way to extensions to more realistic observables \cite[e.g.][]{paolo}. The general formalism involving large deviation theory was first used explicitly for the clustering of the matter field \cite[see for example][]{seminalLDT, saddle}. At roughly the same time as \cite{paolo}, the works of \cite{BernardeauValageas} also inspired \cite{FriedrichDES17} in their modelling of the PDF of tangential shear profiles. The most recent works on the PDF of the convergence and the aperture mass within this large deviation framework are \cite{Barthelemy20a} and \cite{Barthelemy21}, which take into account the full geometry of the light-cone and probe the validity regime of the formalism with the help of numerical simulations.

We organise this paper as follows. In Section 2, we define the CGF and review our theoretical model of weak lensing convergence one-point statistics from LDT. In Section 3, we determine an appropriate sampling for the CGF to minimise the impact of rare events to provide robust cosmological analysis. In Section 4, we examine the covariance matrix of the sampled CGF and perform Fisher forecasts to demonstrate that the PDF and CGF contain equal cosmological information. We provide some discussion in Section 5 and conclude in Section 6. In Appendix A, we validate our estimator for the CGF, and in Appendix B, we discuss the conditioning of the CGF covariance matrix.

\section{Theoretical modelling of the convergence field}
\label{sec::LDT}

\subsection{Useful statistical definitions}

Throughout this work, we make use of different statistical quantities that we briefly introduce here for clarity.
From the PDF, ${\mathcal P}_X$, of some continuous random variable $X$, one can define the moment generating function (MGF) as the Laplace transform of the PDF
\begin{equation}
    M_X(\lambda) =\mathbb{E}\left(e^{\lambda X}\right) =  \int_{-\infty}^{+ \infty} e^{\lambda x} {\mathcal P}_X(x) {\rm d}x,
    \label{laplace}
\end{equation}
or equivalently as the expectation value\footnote{Note that we make use throughout this work of the ergodicity hypothesis, in which one assumes that ensemble averages are equivalent to spatial averages ($\mathbb{E}(.)\rightarrow \left\langle.\right\rangle$) over one realisation of a random field at one fixed time. This requires that spatial correlations decay sufficiently rapidly with separation so that one has access to many statistically independent volumes in one realisation.} of the random variable $e^{\lambda X}$. Note however that the existence of an MGF is not guaranteed for all possible random fields. For example, the MGF of a strictly lognormal field is undefined for real positive $\lambda$. For the MGF to exist (in cases where the PDF also exists), the PDF needs to decay faster than the exponential for the integral in equation~\eqref{laplace} to exist.  In the case of the cosmic density PDF, as computed within the large deviation framework in this paper, there actually exists a critical positive real value $\lambda_c$ -- hereafter dubbed the \textit{critical point} -- beyond which the MGF is not defined. In practice and for a field sampled within a finite volume, the MGF along the real axis will always exist and will simply tend towards $e^{\lambda X_{\rm max}}$ for $\lambda \geq \lambda_c$, where $X_{\rm max}$ is the maximum value of $X$ in the finite field.

The moment generating function, as its name implies, can be used to find the moments of the distribution as can be seen from the series expansion of the expectation of $e^{\lambda X}$,
\begin{equation}
\begin{aligned}
    M_{X}(\lambda)\!&=\!\mathbb{E}\left(e^{\lambda X}\right)\!
    &= \sum_{n = 0}^{+ \infty} \frac{\lambda^{n} \mathbb{E}\left(X^{n}\right)}{n !},
\end{aligned}  
\end{equation}
so that the $n$th derivative of the moment generating function in $\lambda=0$ is equal to the $n$th-order moment, $\mathbb{E}\left(X^{n}\right)$.
The logarithm of the MGF is the CGF 
\begin{equation}
    \phi(\lambda) = \log(M(\lambda)) = \log(\mathbb{E}(e^{\lambda X})) = \sum_{n=1}^{+ \infty} k_{n} \frac{\lambda^{n}}{n !}\,,
    \label{eq:cgf_def} 
\end{equation}
where $k_n$ are the cumulants (i.e. the connected moments) of the distribution and where we have dropped the subscript $X$ for clarity. Note that, in this work, we directly measure the CGF of a given field using this equation. For each value of $\lambda$, we calculate $\exp{(\lambda X)}$ for the entire field and then take the logarithm of the mean.

The reduced cumulants are defined as 
\begin{equation}
    S_n = \frac{k_n}{k_2^{n-1}}, \ n\geq 1
    \label{Sp}
\end{equation}
where $k_2$ is the variance. These are of importance in the context of cosmological structure formation because the $S_n$ of the cosmic matter density field have been shown to be independent of the variance (and therefore redshift) down to mildly non-linear scales \citep{Peebles,1995MNRAS.274.1049B}. We thus also define the scaled cumulant generating function (SCGF hereafter) as
\begin{equation}
    \varphi(\lambda) = \lim_{k_2 \rightarrow 0} \sum_{n=1}^{+\infty} S_n \, \frac{\lambda^n}{n!}=\lim_{k_2 \rightarrow 0} k_2 \,\phi\left(\frac{\lambda}{k_2}\right),
    \label{defscgf}
\end{equation}
which we also extrapolate to non-zero values of the variance (i.e. evaluate at finite $k_2$ on our chosen smoothing scale). One can reconstruct the PDF from the CGF using an inverse Laplace transform (inverting equation~\eqref{laplace})

\begin{equation}
    {\mathcal P}(x) = \int_{-i\infty}^{+i\infty} \frac{{\rm d}\lambda}{2\pi i} \, \text{exp}\left(-\lambda x + \phi(\lambda)\right)\,.
    \label{eq:inverse_laplace}
\end{equation}

\subsection{Large deviation theory of the matter density field}
\label{sec::LDT_def}

The large deviation theory (LDT) framework in large-scale structure has mainly been used to access the one-point statistics of the smoothed density field \citep[see, for example,][]{seminalLDT, saddle, cylindres} and by projection the convergence and aperture-mass fields \citep{Barthelemy20a,Barthelemy21,Boyle21}. It has also been extended to the joint distribution between densities measured at a some distance \citep{2016MNRAS.460.1598C} and projected quantities between different source bins \citep{Barthelemy22}. The results are most simply applied for highly symmetrical window functions such as two- or three-dimensional top hats, but can be generalised to other smoothing schemes \citep{seminalLDT, paolo, Barthelemy21}. We begin here by recalling some of the results of LDT for the one-point statistics of the matter density contrast smoothed in two-dimensional disks (which replicates the dynamics within long cylinders), which in turn will allow us to compute the one-point statistics of projected quantities like the convergence field.

A set of random variables $\{\rho^\epsilon\}_{\epsilon}$ with PDF ${\mathcal P}_{\epsilon}(\rho^\epsilon)$ is said to satisfy a large deviation principle if the limit
\begin{equation}
    \Psi_{\rho^\epsilon}(\rho^\epsilon) = - \lim_{\epsilon \rightarrow 0} \epsilon \log\left[{\mathcal P}_{\epsilon}(\rho^\epsilon)\right]
    \label{LDP}
\end{equation}
exists, where $\epsilon$ is the \textit{driving parameter}. $\Psi$ is known as the rate function of $\rho^\epsilon$ and describes the exponential decay of its PDF. The driving parameter $\epsilon$ indexes the random variable with respect to some evolution, for example an evolution in time. 
In the case of the matter density field smoothed on a single scale $R$, this driving parameter is the variance, which acts as a clock from initial to late times ($\epsilon \equiv \sigma^2_R$). We now omit the $\epsilon$ sub/superscripts in our notation for simplicity.

The existence of a large deviation principle for the random variable $\rho$ implies that its SCGF $\varphi_{\rho}$ is given through Varadhan's theorem as the Legendre-Fenchel transform of the rate function $\Psi_{\rho}$ \citep{Ellis_book,touchette}
\begin{equation}
    \varphi_{\rho}(\lambda) = \sup_{\rho} \,\left[ \lambda\rho - \Psi_{\rho}(\rho)\right],
    \label{varadhan}
\end{equation}
where the Legendre-Fenchel transform reduces to a simple Legendre transform when $\Psi_{\rho}$ is convex. In that case, 
\begin{equation}
    \varphi_{\rho}(\lambda) =  \lambda\rho - \Psi_{\rho}(\rho),
    \label{Legendre}
\end{equation}
where $\rho$ is a function of $\lambda$ through the stationary condition \footnote{The $\rho_c$ value at which $\Psi_{\rho}$ ceases to be convex leads to a $\lambda_c$ value that corresponds to the critical point mentioned in the discussion of equation~\eqref{laplace}.} 
\begin{equation}
   \lambda = \frac{\partial \Psi_{\rho}(\rho)}{\partial \rho}.
    \label{stationnary}
\end{equation}
Another consequence of the large-deviation principle is the so-called contraction principle.
This principle states that for a random variable $\tau$ satisfying a large deviation principle and related to $\rho$ through the continuous mapping $f$, the rate function of $\rho$ can be computed as
\begin{equation}
    \Psi_{\rho}(\rho) = \inf_{\tau:f(\tau) = \rho} \Psi_{\tau}(\tau).
    \label{contraction}
\end{equation}
This is called the contraction principle because $f$ can be many-to-one, in which case we are {\it contracting} information about the rate function of one random variable down to the other. In physical terms, this states that an improbable fluctuation of $\rho$ is brought about by the most probable of all improbable fluctuations of $\tau$.

For the case of the matter density field, the rate function of the late-time density field at different scales can be computed from the initial conditions if the most likely mapping between the two is known -- that is, if one is able to identify the leading field configuration that will contribute to the infimum of equation~\eqref{contraction}. In cylindrically symmetric configurations, as for a disk of radius $R$ at redshift $z$ (in 2D space) or alternatively a very long 3D cylinder centered on this disk,
the most likely mapping between final and initial conditions should preserve the symmetry\footnote{This is only true for a certain range of density contrasts around zero, very much sufficient for our purposes. However, one could note that there are counter-examples in which the spherical or cylindrical symmetry does not lead to spherical/cylindrical collapse being the most likely dynamics, for example in 1D for very high values of the density.} \citep{Valageas},
 which leads to initial conditions also being cylindrically symmetric and the dynamics between the two being those of cylindrical collapse.

Thus, starting from Gaussian initial conditions,\footnote{Primordial non-Gaussianities can also straightforwardly be accounted for in this formalism as shown by \cite{NonGaussianities}.} the rate function of the linear field is simply given by a quadratic term. The rate function of the late-time density field in a disk of radius $R$ is then given by
\begin{equation}
    \Psi_{\rm cyl}(\rho)=\sigma^2_{R}\frac{\bar{\tau}^2}{2 \sigma^2_{r}},
    \label{psicyl}
\end{equation}
where $\sigma^2_{R}$ --- our driving parameter --- is the variance of the non-linear density field in the disk, $\sigma^2_{r}$ is the variance of the linear density field inside the initial disk (before collapse) of radius $ r = R \, \rho^{1/2}$ (from mass conservation), and $\bar{\tau}$ is the linear density contrast obtained through the most probable mapping between the linear and late-time density fields. This mapping is given by 2D spherical (cylindrical) collapse, for which an accurate parametrisation is given by \citep{Bernardeau1995}
\begin{equation}
    \zeta(\bar{\tau}) = \rho =  \left(1 - \frac{\bar{\tau}}{\nu} \right)^{-\nu}.
    \label{collapse}
\end{equation}
In the spirit of previous works involving the density filtered in spherical cells, the value of $\nu$ in this parametrisation of $\zeta$ is chosen to be $\nu = 1.4$, so as to reproduce the value of the tree-order skewness in cylinders as computed from perturbation theory \citep{cylindres}.

Then, as a straightforward consequence of the contraction principle, the rate function given by equation~\eqref{psicyl} is also the rate function of any monotonic transformation of $\rho$, so that for the density contrast $\delta = \rho - 1$, we have $\Psi_{\delta}(\delta) = \Psi_{\rho}(\rho(\delta))$. Thus, plugging equation~\eqref{psicyl} into equation~\eqref{Legendre} gives us the SCGF of the matter density contrast in a disk at redshift $z$. 

Finally, one of the key aspects of the large deviation formalism in the cosmological context is that we apply the result for the SCGF beyond the $\sigma^2_R \rightarrow 0$ limit. This allows us to extrapolate to the CGF of the real density field by rescaling the SCGF by the driving parameter (the non-linear variance) at the scale and redshift being considered, such that
\begin{equation}
    \phi_{\rm cyl}(\lambda) = \frac{1}{\sigma^2_R}\varphi_{\rm cyl}(\lambda \sigma^2_R). 
\end{equation}
This is physically meaningful because the reduced cumulants $S_n$ from the cylindrical/spherical collapse dynamics have been shown to be very robust over a large range of scales and redshifts down to mildly non-linear scales ($\gtrsim 5$ Mpc/$h$ at $z \gtrsim 0$ (see for example Figure~2 in \cite{saddle} and Figure A1 in \cite{cylindres}), so that re-scaling by the non-linear variance allows access to the full one-point statistics of the non-linear density field. In LDT terms, the SCGF given by the large deviation principle is the well-defined asymptotic form taken by the cumulants of the field in the regime where the variance goes to zero, and we simply keep this form for our predictions of the non-linear CGF. 

Finally, note that although equations~\eqref{Legendre} and \eqref{psicyl} have been known and used for three decades in the context of counts-in-cells statistics \citep[that is, the density field filtered in 3D top-hat windows --- see for example][]{Bernardeau_1994a,Valageas_1998}, their re-derivation through large deviation statistics is more general and allows one to set up a framework for the computation of different probabilities in the cosmological context. 

\subsection{From matter density to convergence}

Let us recall that for a flat cosmology, the convergence $\kappa$ can be interpreted as a line-of-sight projection of the matter density distribution between the observer and the source and can be written as \citep{kappadef}
\begin{equation}
    \kappa({\bm \vartheta}) = \int_0^{z_s} {\rm d}z \frac{{\rm d}\chi}{{\rm d}z} \, \omega(\chi,\chi_s) \, \delta(\chi,\chi{\bm \vartheta}),
    \label{def-convergence}    
\end{equation}
where $\chi$ is the comoving radial distance, $z_s$ is the source redshift, and $\chi_s$ is the cosmology-dependent distance associated with the source redshift. 
The lensing kernel $\omega$ is then defined as 
\begin{equation}
\label{eq:weight}
    \omega(\chi,\chi_s) = \frac{3\,\Omega_m\,H_0^2}{2\,c^2} \, \frac{\chi(\chi_s-\chi)}{\chi_s}\,(1+z(\chi)).
\end{equation}

Under the small-angle/Limber approximation, it can be shown that correlators of the (smoothed) convergence field can be seen as a juxtaposition of the 2D correlators of the underlying density field, as if each 2D slice along the line of sight is statistically independent of the others \citep{BernardeauValageas,Barthelemy20a}. In terms of the one-point statistics of the smoothed $\kappa$ field within a top-hat window function of angular radius $\theta$, this translates to saying that the CGF of $\kappa$ is a sum along the line of sight of the CGF of independent 2D slices 
of the matter density contrast:\footnote{Rigorously, this result applies for a juxtaposition of very long cylinders centered on the slices and of length $L\rightarrow
\infty$. Since the symmetry and thus the most likely dynamics of these long cylinders are the same as for a 2D slice in a 2D space, and since the results are independent of $L$, we refer to \lq 2D slices\rq~for clarity. This emphasises that correlations along the line of sight are negligible compared to those in the transverse directions.}
\begin{equation}
    \phi_{\kappa,\theta}(\lambda) = \int_0^{\chi_s} {\rm d}\chi \, \phi_{{\rm cyl}}^{<\chi \theta}(\omega(\chi,\chi_s) \lambda,\chi(z)),
    \label{projection}
\end{equation}
where $\phi_{{\rm cyl}}^{<\chi \theta}$ is the CGF of the density contrast filtered within a disk 
of radius $\chi(z) \theta$ so as to reproduce the geometry of the light cone.

Equation~\eqref{projection} thus reduces the complexity of the problem down to computing the one-point statistics of the 2D matter density in each two-dimensional slice (or equivalently 
within long 3D cylinders at the same redshift) along the line of sight, which we have already done in Section~\ref{sec::LDT_def}. Using these results, we can then build the non-linear CGF of the convergence field. Note that equation~\eqref{projection} highlights the nice property of the projected CGF being expressible simply as a sum of independent redshift slices. This is an important property when considering more complicated joint distributions, in which the only modification in this integral would be the replacement of the $\omega(\chi,\chi_s)\lambda$ term by a term depending on more than one $\lambda$ variable. This form is much simpler than that of the corresponding multi-variate PDF \citep{Barthelemy22}.

Finally, it is important when working with the CGF to know the approximate location of the (theoretical) critical point of the convergence field, $\lambda_c$. First, the critical points $\delta_c$ in each redshift slice are calculated by finding where the second derivative of the rate function becomes equal to zero. The corresponding critical $\lambda$ values can be obtained by applying the stationary condition (Eq. \eqref{stationnary}). The minimum $\lambda_c$ along the line of sight is taken as the critical point of the convergence CGF.


\section{The Cumulant Generating Function as an Alternative Observable}

\begin{figure}
    \centering
    \includegraphics[width = \columnwidth]{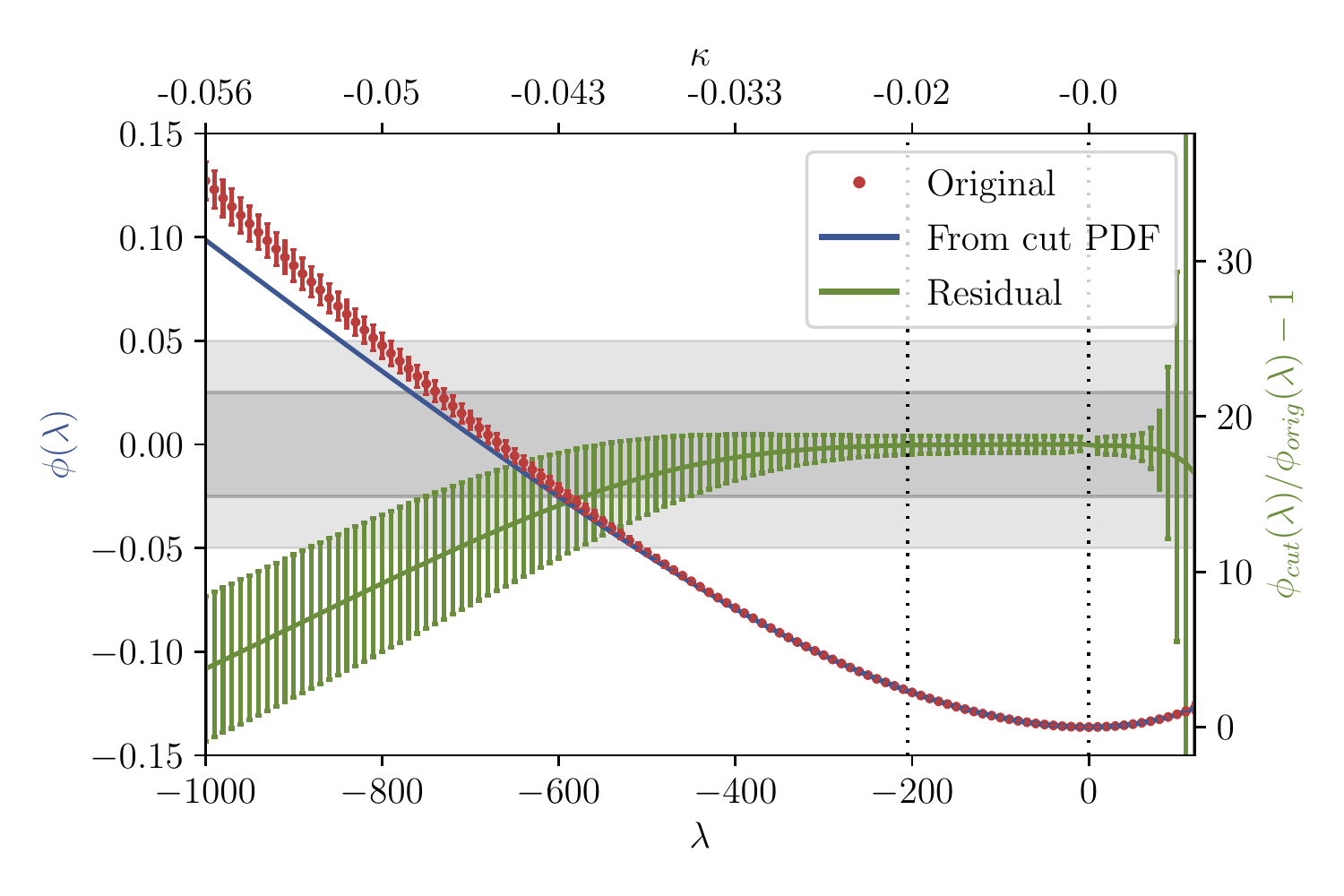}
    \caption{
    The impact of small tail cuts in $\kappa$ on the corresponding CGF ($z_s=2$, $\theta=10$ arcmin). The red line shows the CGF calculated from the large deviation formalism, with the imposed error bars measured from the Takahashi simulation convergence maps. The blue line shows the CGF obtained from a Laplace transform of the PDF cut to exclude $\kappa$ values with probability densities lower than $10^{-3}$ ($\kappa \in [-0.045;0.07]$). The green line shows the residual between the CGF obtained from the cut CGF and the original. This roughly shows what happens on a simulated small convergence map where rare events are not present, and demonstrates that using either the CGF or PDF as an observable should yield similar information content, because the largest deviations are seen far in the negative $\lambda$ tail, which we will see is not needed to extract cosmological information. The vertical dotted lines indicate the section of the CGF we eventually use for our Fisher analysis.} 
    \label{CGF_from_cut_PDF}
\end{figure}


Although the large deviations formalism presented in Section~\ref{sec::LDT} initially provides predictions for the full CGF, all previous work has focused on the corresponding PDF (as a complicated inverse Laplace transform \eqref{eq:inverse_laplace}) or on individual cumulants (as Taylor coefficients of the CGF~\eqref{eq:cgf_def}). The main reason for this seems to be the intuitiveness of these quantities --- the PDF is easily interpreted as a simple histogram and the cumulants are visualisable as properties of the shape of the PDF. 
However, the CGF is clearly also an observable (the log of the expectation value in equation~\eqref{eq:cgf_def}).

Beyond simplicity, another motivation for concerning ourselves with the CGF instead of the PDF is as follows. In recent work, \cite{Barthelemy22} explored the computational complexity involved in calculating the joint PDF for a tomographic weak lensing survey.
Despite some simplification strategies, an exact calculation of the joint PDF remains computationally expensive, with the computational time required increasing rapidly with the number of source redshift bins considered. However, this complexity mostly comes from the fact that the correlation structure between bins is expressed in a very complicated manner for the PDF, while it is much more transparent for the CGF. As a consequence, working with the CGF directly should significantly improve the efficiency of our calculations. This is particularly important for cosmological forecasts and analyses, where it is desirable to be able to rapidly produce a large number of predictions for different sets of cosmological parameters. Additionally, note that although \cite{Barthelemy22} focused explicitly on joint-lensing statistics between redshift bins, their result could be 
summarised by stating that the Limber approximation allows drastic simplifications of the correlation structure between CGFs, and generalised to the case of the (fully projected or tomographic) density field at different redshifts. This could be applied to joint analyses of lensing and clustering, making the inclusion of their full joint distributions more feasible.

For the CGF to be a robust observable for cosmological forecasting, it has to be demonstrated that it changes minimally when very rare events are excluded, in order to allow for measurement limitations and small survey sizes. Figure~\ref{CGF_from_cut_PDF} compares a CGF calculated directly from the LDT formalism (red) with one Laplace-transformed from a truncated PDF (blue), with the rare-event tails cut out to exclude values of $\kappa$ with probability densities below $10^{-3}$. The result is very satisfactory, with very little difference seen between the two. The residual is shown in green, with error bars derived from 108 convergence maps from the Takahashi simulations (see Section~\ref{sec:measuring_the_cgf_in_simulations}). The maximal difference of a few percent occurs at very low $\lambda$ values that we will see later are not needed to extract cosmological information. 
Note that this difference at low $\lambda$ is expected, because the asymptotic behaviour of the negative tail of the CGF is $\propto \lambda \kappa_{\rm min}$, where $\kappa_{\rm min}$ is either the minimum theoretical value the convergence can take (when the density contrast equals -1 all along the line of sight), or the minimum value of $\kappa$ after the cut or in a small simulated field.

\subsection{Relating $\kappa$ and $\lambda$ with the saddle-point approximation}
\label{kappa-lambda-relation}

One primary goal of this work is to demonstrate that for practical purposes the PDF and CGF contain equal information, by determining how to accurately translate information content between the two. While the PDF is generally considered in $\kappa$ bins of (usually equal) finite width, the CGF is sampled at individual $\lambda$ values. We can apply a saddle-point approximation to the inverse Laplace transform in equation~\eqref{eq:inverse_laplace} such that
\begin{figure}
    \centering
    \includegraphics[width=\columnwidth]{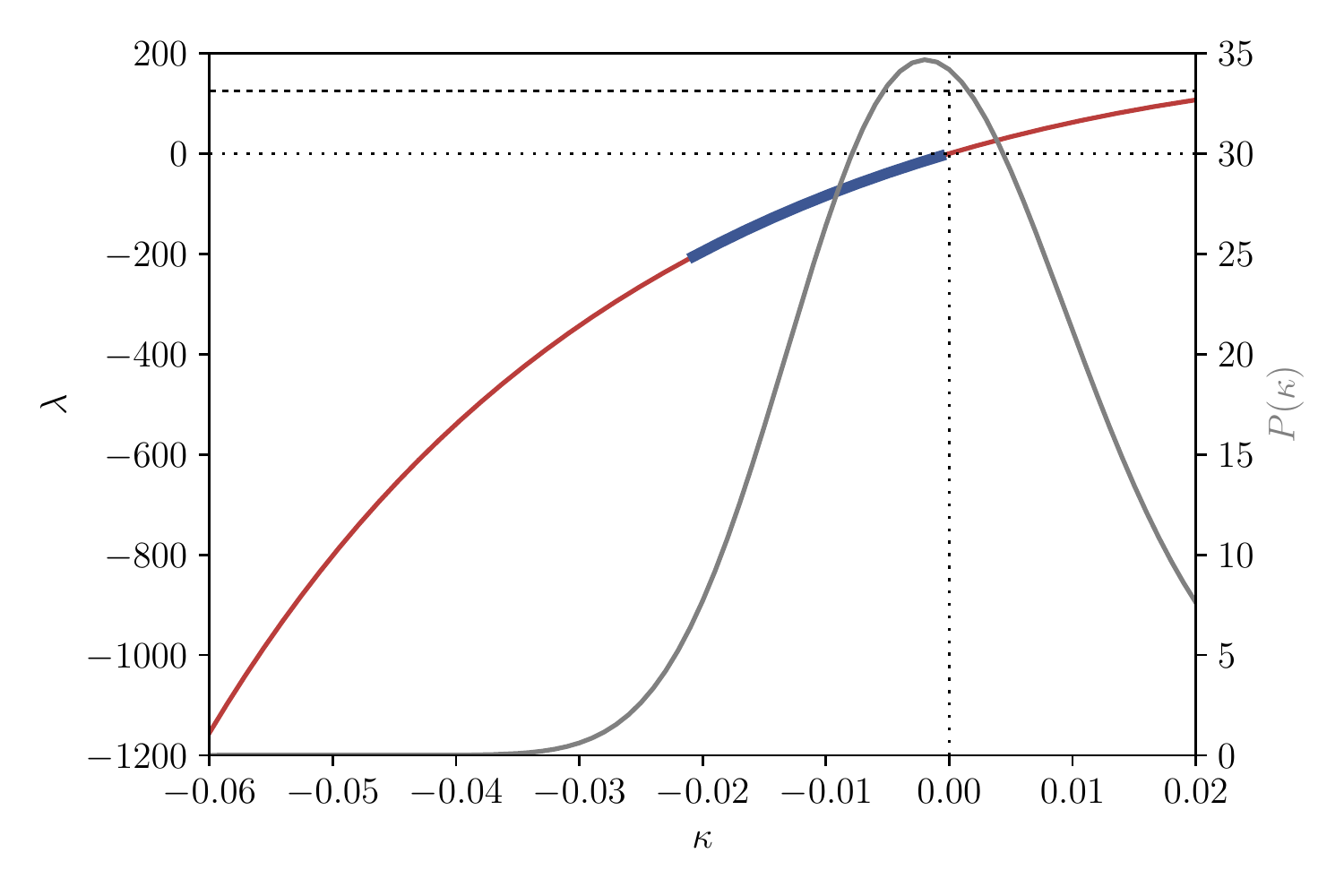}
    \caption{
    The relationship between values of $\kappa$ and $\lambda$ inferred from the saddle-point approximation \eqref{eq:PDF_saddle}. The blue section shows the region of $\lambda$ values that we use in practice in our Fisher forecasts, covering the range from $\kappa_{\rm min}$ to zero. As explained later, we find that using only the negative part of the CGF is sufficient to reproduce the constraints from the bulk of the PDF. The denser dotted line at $\lambda=125$ represents the theoretical critical point. The theoretical PDF is shown in the background in grey to provide intuition for the corresponding abundance of $\kappa$ values.}
    \label{fig:kappa_lambda}
\end{figure}
\begin{equation}
\label{eq:PDF_saddle}
    \mathcal P(\kappa)\approx\frac{1}{\sqrt{2\pi}}\sqrt{\frac{\partial^2 \psi(\kappa)}{\partial\kappa^2}}\exp(-\psi(\kappa)),
\end{equation}
where $\phi_{\kappa}(\lambda)=-\psi(\kappa)+\lambda\kappa$ with the stationary condition $\lambda=\partial\psi(\kappa)/\partial \kappa$. This gives an approximate direct relationship between values of $\kappa$ and $\lambda$, which we use as a guideline for predicting appropriate values of $\lambda$ to sample to obtain approximately the same information from the CGF as from a given PDF. This relationship is illustrated in Figure~\ref{fig:kappa_lambda}. The blue section highlights the range of $\lambda$ values we ultimately use in our Fisher forecasts (corresponding to $\kappa$ values between the $\kappa_{
\rm min}$ we adopt for the PDF and zero). The choice to use zero as the upper limit in $\lambda$ instead of a $\lambda$ value derived from $\kappa_{\rm max}$ is explained in Section \ref{subsec::GaussianLikelyhood}, while its impact in terms of cosmological information is discussed in Section \ref{subsec::negativeLambda}.





\subsection{Measuring the CGF in Simulations}\label{sec:measuring_the_cgf_in_simulations}

\begin{figure}
    \centering
    \includegraphics[width=\columnwidth]{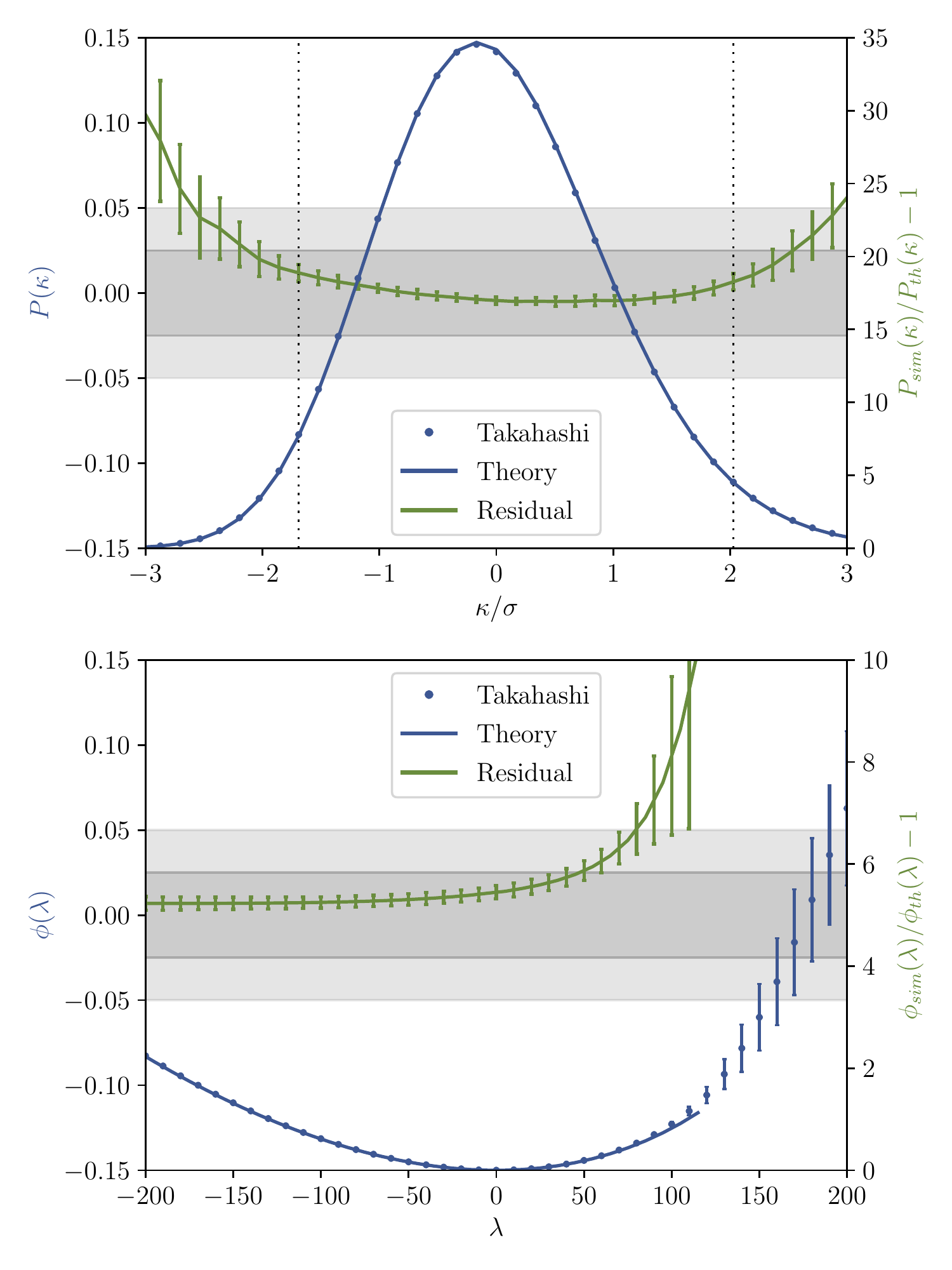}
    \caption{\textbf{Upper panel:} A comparison of the theoretically modelled PDF and that from the Takahashi simulations (for fixed source redshift $z_s\approx 2$ and a smoothing scale of 10 arcmin). The theoretical and simulated PDFs are shown in blue (see the left y-axis). The residual is shown in green (see the right y-axis) and shows percent-level accuracy within the region considered. The vertical dotted lines indicate the portion of the PDF used for the Fisher matrix forecast (which is decided based on a $\chi^2$ test to ensure the Gaussianity of the data vector). The horizontal grey bands represent a 2.5\% and 5\% deviation. \textbf{Lower panel:} A comparison of the corresponding CGFs. The theoretical and simulated CGFs are shown in blue (see the left y-axis). The theoretical CGF is shown up to $\lambda=125$, which approximately corresponds to the critical point beyond which the CGF develops linear asymptotic behaviour (evolving $\propto\lambda\kappa_{\rm max}$) with rapidly increasing error bars. This behaviour is shown as measured from the maps by the blue points. The error bars on the residual become larger around $\lambda=100$, indicating that the critical value in the measured CGF is lower than that predicted by the theory, as expected. The residual between the theoretical and measured CGFs is shown in green (see the right y-axis).}
    \label{fig:pdf_cgf}
\end{figure}

To validate our model for the CGF and to generate a covariance matrix, we use a set of weak lensing convergence maps generated from ray-traced N-body simulations as part of the Takahashi simulation suite \citep{takahashi_full-sky_2017}. We have 108 maps for a fixed source redshift of $z_s\approx 2$. We smooth the maps with a top-hat filter of radius 10 arcmin in harmonic space. We extract the PDF in 50 linearly-spaced bins in the range $-0.1\leq\kappa\leq 0.1$ and the CGF for real values of $\lambda$ in the range $-1000\leq\lambda\leq 150$ with a spacing of 5, having calculated that the theoretical critical point in the CGF lies at approximately $\lambda=125$. The CGF is calculated directly as we define it in equation~\eqref{eq:cgf_def} -- for each value of $\lambda$, we take the logarithm of the expectation value of $\exp(\lambda \kappa)$, it is neither reconstructed from the PDF nor from cumulants. Figure~\ref{fig:pdf_cgf} shows the theoretical and measured PDFs (upper panel) alongside the residual between them in green; the corresponding result for the CGF is shown in the lower panel. We also demonstrate in Appendix~\ref{app:estimator} that our estimator for the CGF is robust and discuss its validity and potential biases.

\section{Cosmological Analysis}\label{sec:cosmological_analysis}

\subsection{Covariance matrix}

\begin{figure}
    \centering
    \includegraphics[width=\columnwidth]{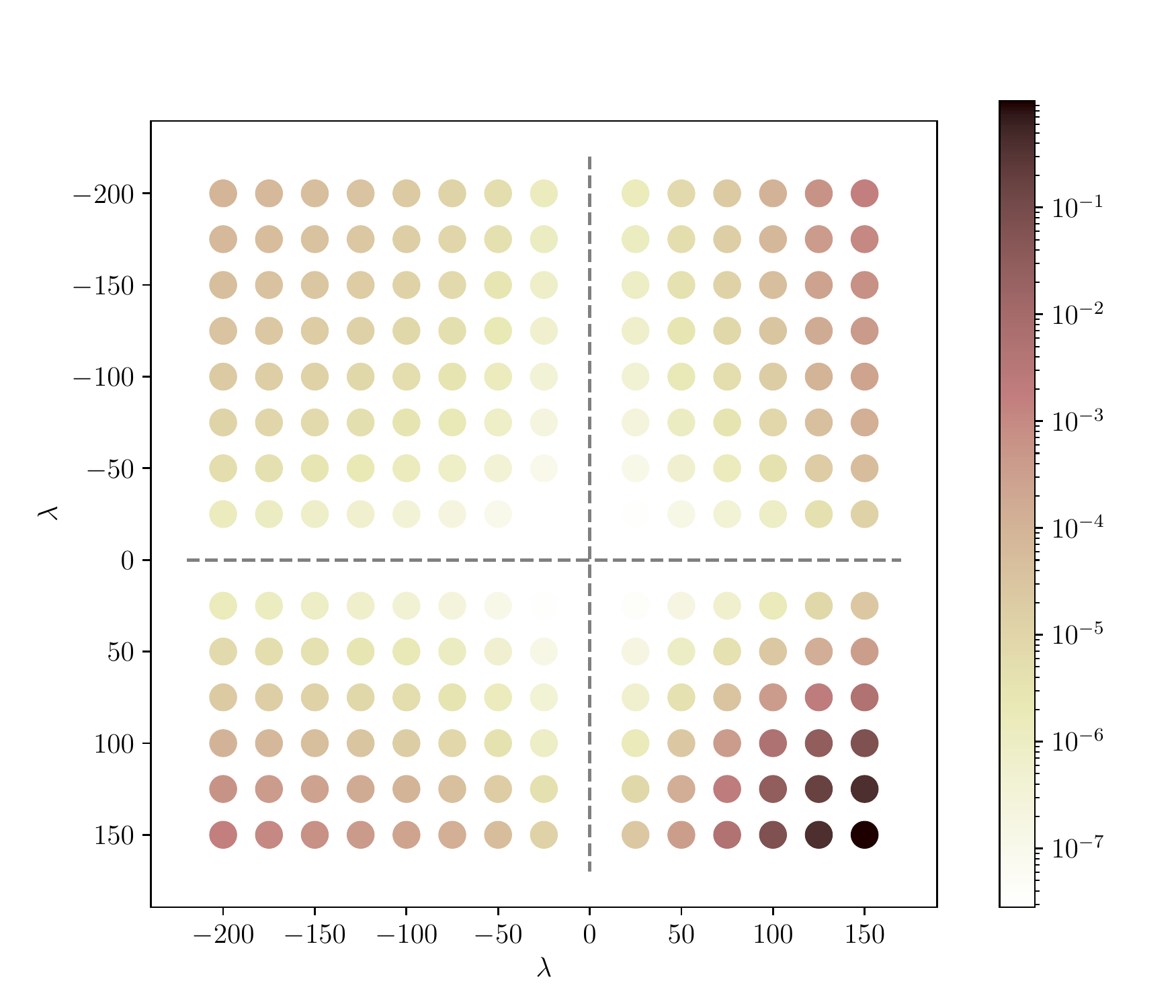}
    \caption{The covariance matrix of the CGF as measured for a sample of $\lambda$ values. The lower $\lambda$ value of -200 is extracted from the lowest $\kappa$ value used for the PDF using the saddle-point approximation \eqref{eq:PDF_saddle}. The critical point is at approximately 125 for the theoretical CGF but is lower in the actual maps. We can observe a number of distinctive features. First, the values of $\phi(\lambda)$ are all positively correlated with each other (note that the colour bar only covers correlation coefficients between zero and one and has a logarithmic scale). Second, the large covariance values for higher $\lambda$ values correspond to the expected behaviour of errors beyond the critical point.}
    \label{fig:covariance}
\end{figure}

Figure~\ref{fig:covariance} shows the covariance matrix of the CGF as measured from 108 Takahashi simulation convergence maps. The lower $\lambda$ limit is determined from the saddle-point approximation \eqref{eq:PDF_saddle} and our corresponding lower $\kappa$ limit for the PDF (see Figure~\ref{fig:pdf_cgf} for precise values), and the upper limit goes a little beyond the critical point. Note that the colour bar range includes only positive values (on a logarithmic scale) because the points in the CGF are always positively correlated with each other. This can be understood intuitively as a consequence of the positive correlations between the cumulants (due to their dependence on the variance) and the polynomial form of the CGF seen in equation \eqref{eq:cgf_def}. We see a divergence in the covariance for high $\lambda$ values beyond the critical point, as we expect, and which corresponds to the large error bars beyond the critical point seen in Figure~\ref{fig:pdf_cgf}.

\begin{figure}
    \centering
    \includegraphics[width=\columnwidth]{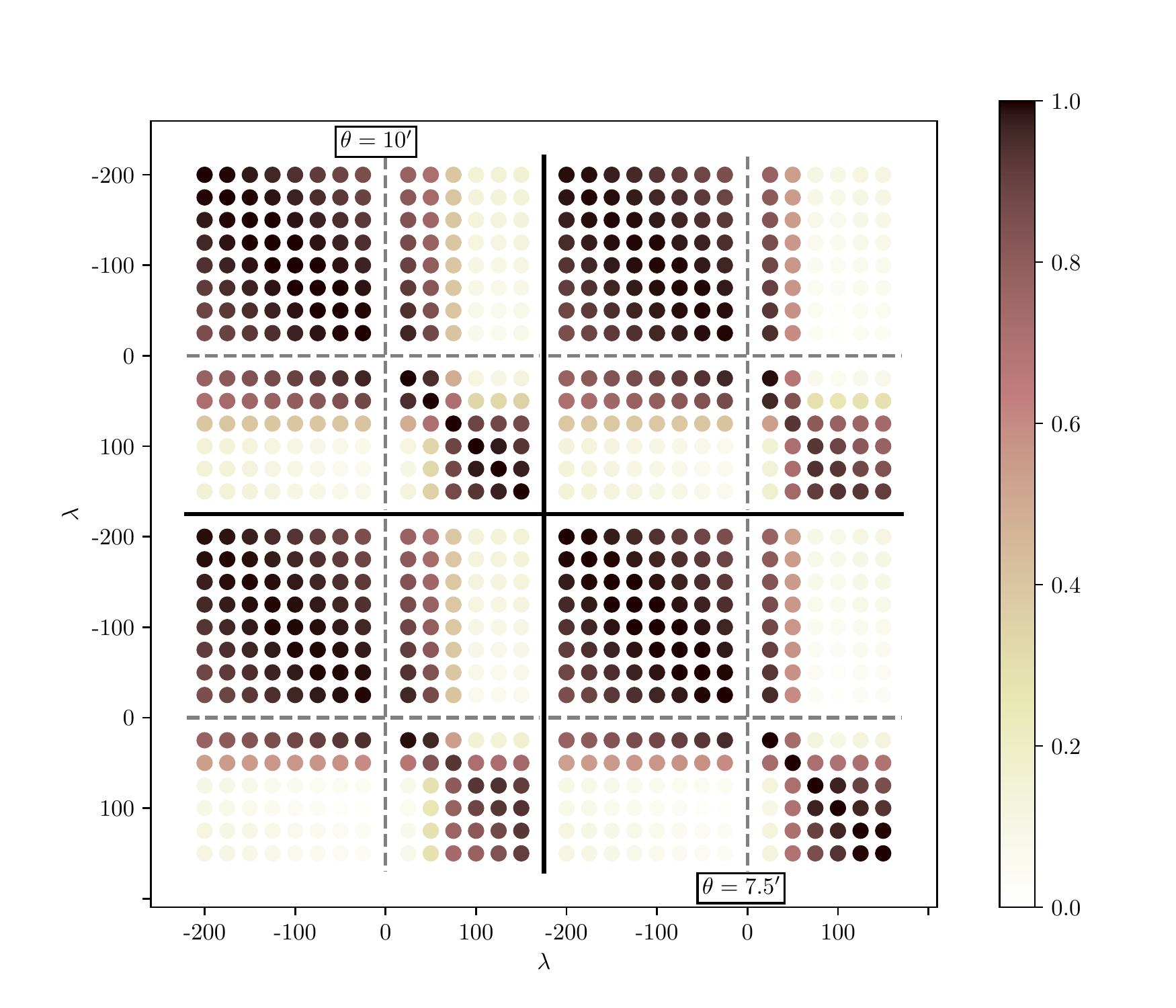}
    \caption{The \textit{correlation} matrix for CGFs on two smoothing scales: 10 arcmin on the top left and 7.5 arcmin on the bottom right, with the cross-correlation between the two scales shown in the two other panels. We note that the top-left panel corresponds to the case shown in the covariance matrix in Figure~\ref{fig:covariance}. We emphasise the very strong correlation between the negative-$\lambda$ CGF points and between the two CGF smoothing scales. We also note the very strong correlation between points beyond the critical point and the very weak correlation between these points and the rest of the CGF --- both behaviours are expected because these points are sensitive only to the maximum values of $\kappa$ in the individual maps.}
    \label{fig:covariance_2scale}\label{fig:correlation_matrix}
\end{figure}

In Figure~\ref{fig:covariance_2scale} we show a \textit{correlation} matrix for a data vector that combines CGFs for two smoothing scales ($\theta=7.5^\prime$ and $\theta=10^\prime$).\footnote{Note that in previous work \citep{Boyle21}, it was shown that combining PDFs with different smoothing scales provided a significant enhancement in cosmological information.} While the covariance matrix is used in Fisher matrix forecasts, the correlation matrix has the advantage of being more visually intuitive for quantities of different magnitudes. The upper left panel corresponds to the covariance matrix seen in Figure~\ref{fig:covariance_2scale}. We see that the two CGFs are also very strongly correlated with each other, as is the case for the corresponding PDFs. We also note that beyond the critical point, the CGF values are only weakly correlated with those before the critical point but very strongly correlated with each other. This is as we expect, because the behaviour of the CGF here is determined only by the maximum $\kappa$ value in a particular map. We note that the critical point in the maps occurs at lower $\lambda$ (around 100, though a precise estimation does not exactly make sense in a finite volume) than in the theory (where we predict $\lambda_{\rm crit}=125$).

\subsection{Ensuring a Gaussian-distributed data vector}
\label{subsec::GaussianLikelyhood}

Fisher matrix forecasts rely on the assumption that the observable data vector is Gaussian distributed. We now need a robust test to determine whether this is the case for the CGF, and indeed whether it is the case for all values of $\lambda$. The $\chi^2$ test is a very useful tool for this purpose, as was demonstrated, for example, in \cite{friedrich_2021}. If the measurements of the elements of a data vector $\bm{s}$, with $\hat{{s}}_i$ one particular measurement, are Gaussian distributed, then the distribution of values of $
\chi^2_i$ defined as
\begin{equation}\label{eq:chi2}
    \chi^2_i = (\hat{s}_i-\mu)C^{-1}(\hat{s}_i-\mu),
\end{equation}
where $\mu$ is the mean measured data vector, should follow a $\chi^2$ distribution for the appropriate number of degrees of freedom (the number of elements in the data vector). Next, one can draw random Gaussian data vectors from the covariance matrix and use them in equation~\eqref{eq:chi2} to see if these follow a similar distribution. If so, we are assured that the data vector considered is reasonably Gaussian-distributed.

The above approach limits us to 108 samples for the inferred distribution, which can be a little limiting when trying to determine how well the result matches a $\chi^2$. As an alternative, one can combine pairs of measurements to increase the number of samples to $108\cdot 107/2$ while preserving the covariance, as
\begin{equation}
    \chi^2_{ij}=\frac{1}{2}(\hat{s}_i-\hat{s}_j)C^{-1}(\hat{s}_i-\hat{s}_j).
\end{equation}
This approach is adopted here and leads to a visually better result for a limited number of measurements. 

\begin{figure}
    \centering
    \includegraphics[width=\columnwidth]{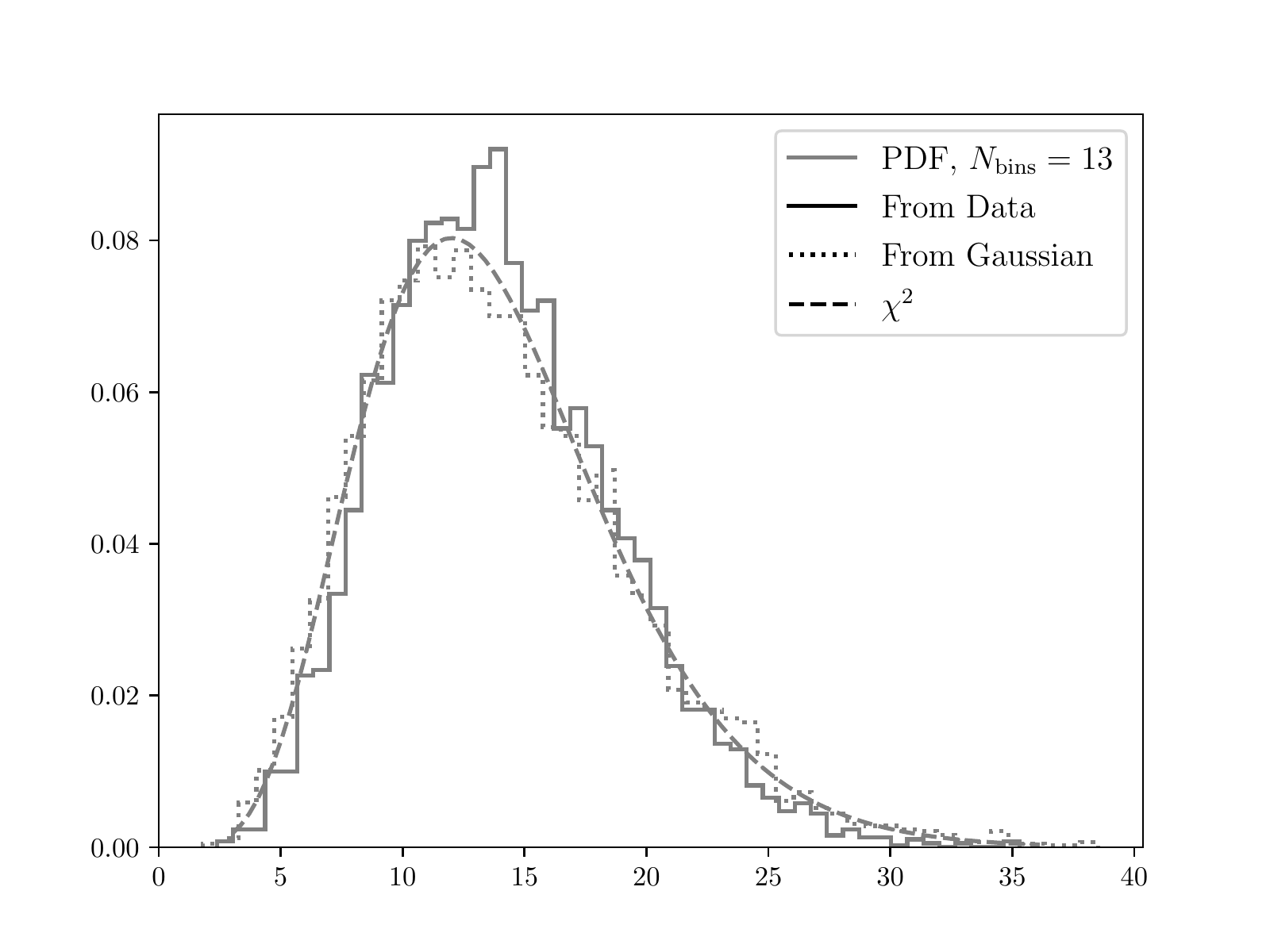}
    \caption{$\chi^2$ test for the PDF. There are 13 $\kappa$ bins between $\kappa_{\rm min}=-0.02$ and $\kappa_{\rm max}=0.026$. The $\chi^2$ test performs well, indicating that the data vector is Gaussian-distributed.}
    \label{fig:chi2_pdf}
\end{figure}

Figure~\ref{fig:chi2_pdf} shows a sample $\chi^2$ test for the PDF, including 13 $\kappa$ bins between $\kappa_{\rm min}=-0.02$ and $\kappa_{\rm max}=0.026$. This is a relatively small number of bins, but is chosen to keep the $\chi^2$ test reliable as our covariance matrix is based on a rather small number of convergence maps (108 from the Takahashi simulations). \cite{Uhlemann2022} showed that a robust covariance matrix can be extracted for twice the number of bins in the same $\kappa$ range with 500 maps. We see that our result coincides reasonably well with a $\chi^2$ distribution for 13 degrees of freedom. 

\begin{figure}
    \centering
    \includegraphics[width=\columnwidth]{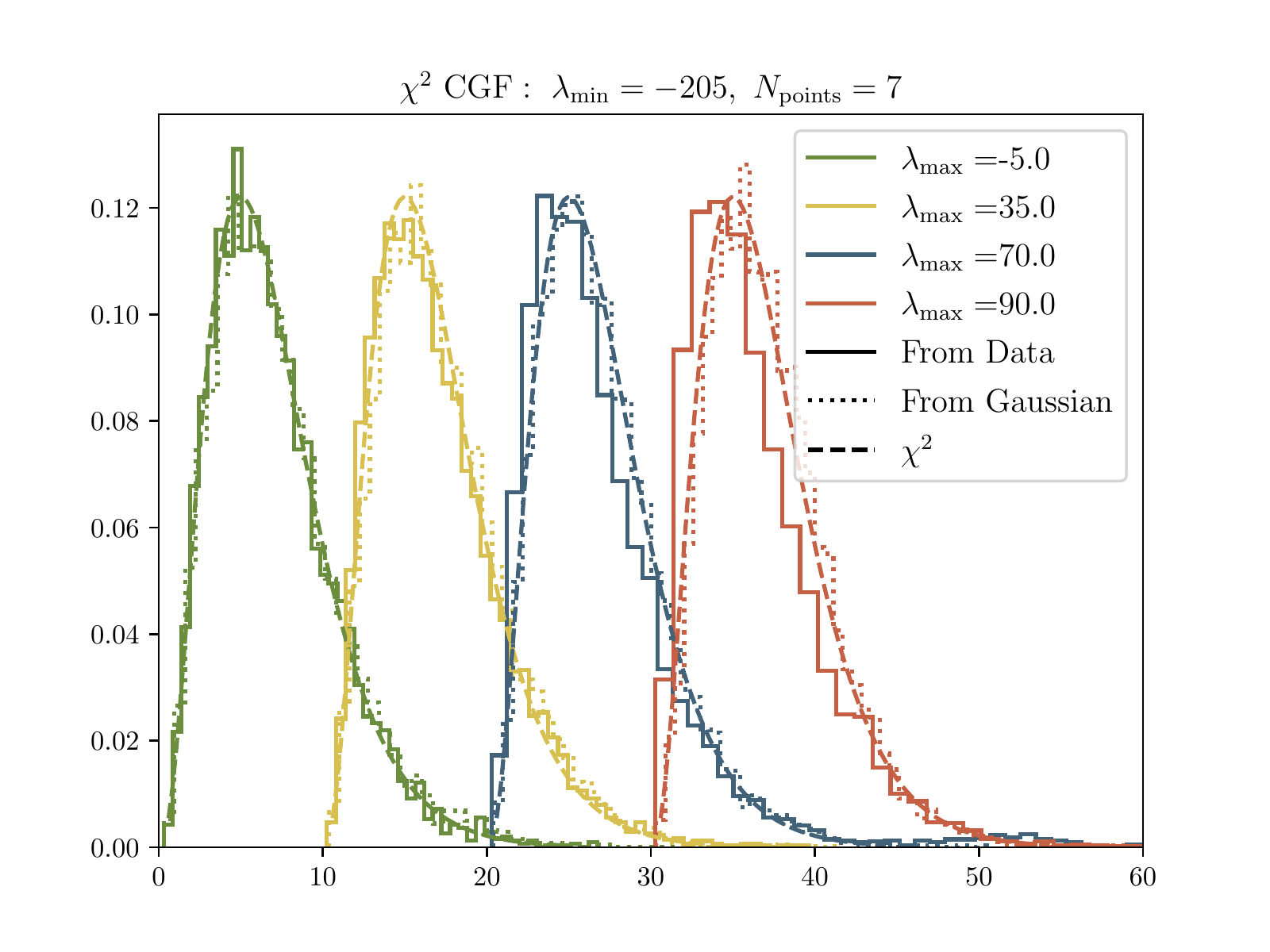}
    \caption{$\chi^2$ tests for the CGF. The different instances are shifted rightwards by an arbitrary amount for clearer viewing. The lowest value of $\lambda$ sampled (-205) is determined from the minimum value of $\kappa$ used in the PDF analysis using the saddle-point approximation. The success of the $\chi^2$ test seems relatively insensitive to the choice of $\lambda_{\rm min}$. The maximum value of $\lambda$ is varied with the number of points sampled being fixed. The $\chi^2$ test is less successful as $\lambda_{\rm max}$ is pushed to higher values and closer to the critical point.}
    \label{fig:chi2_cgf_kappa_maxcuts}
\end{figure}

\begin{figure}
    \centering
    \includegraphics[width=\columnwidth]{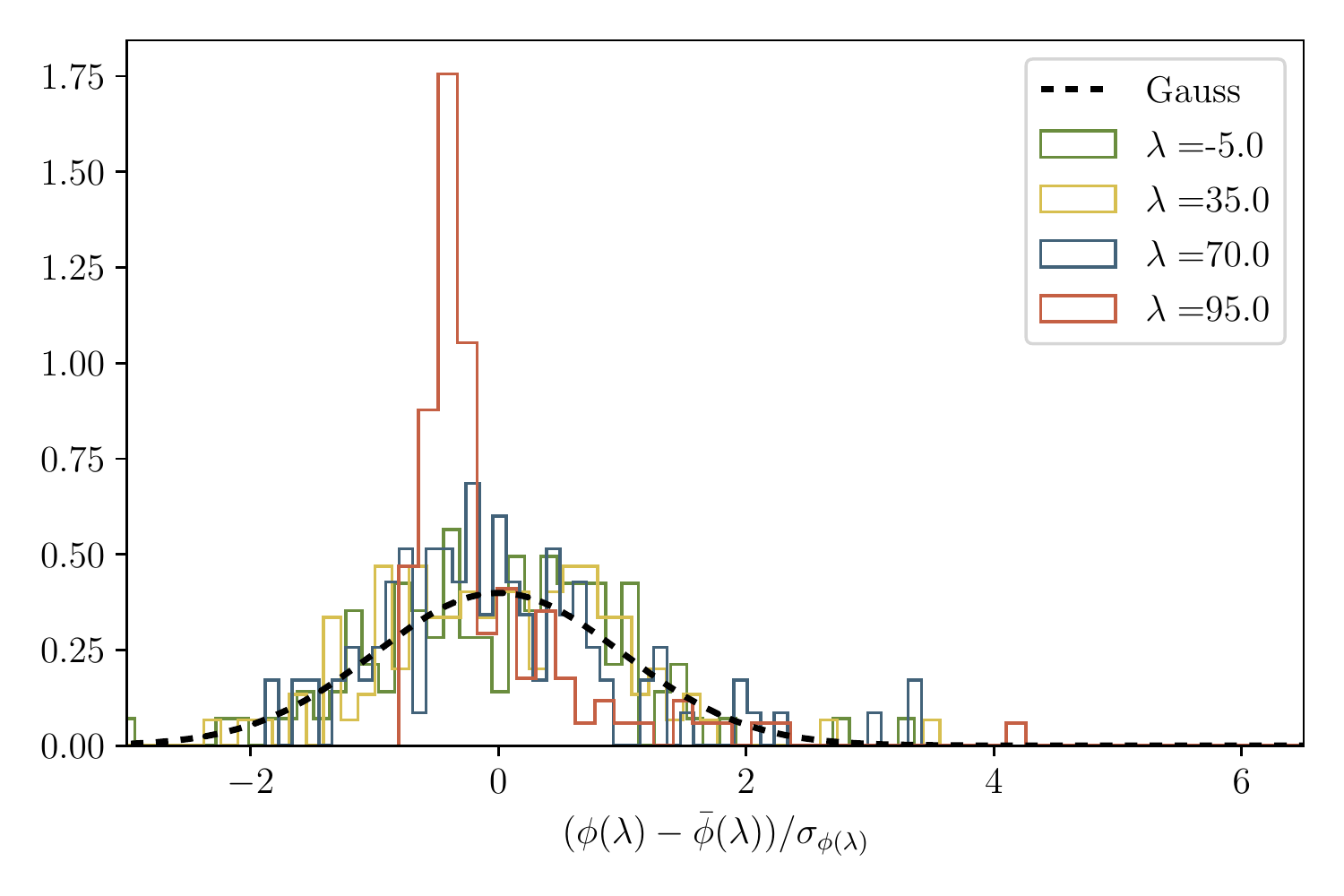}
    \caption{Distributions of $\phi(\lambda)$ for the $\lambda$ values corresponding to the maximum values in Figure~\ref{fig:chi2_cgf_kappa_maxcuts}. The distributions become more and more non-Gaussian on approaching the critical point, with a very clear non-Gaussianity for $\lambda=95$, which explains the result seen in Figure~\ref{fig:chi2_cgf_kappa_maxcuts}.
    }
    \label{fig:cgf_distributions}
\end{figure}

Figure~\ref{fig:chi2_cgf_kappa_maxcuts} shows a number of $\chi^2$ tests for the CGF. The lower value of $\lambda$ is fixed at $-205$ (as inferred from the saddle point approximation and the value of $\kappa_{\rm min}$ used for the PDF) and various values are chosen for the upper limit $\lambda_{\rm max}$, with $\lambda_{\rm max}$ increasing on moving from left to right. Note that $\lambda=-5$ is the last negative $\lambda$ point we can sample because we sample $\lambda$ with a spacing of 5, and we never include $\lambda=0$ because its value is fixed at $\phi(0)=0$ by definition. We see that including points close to the critical point significantly degrades the performance of the $\chi^2$ test. This is probably because the behaviour of the CGF at high $\lambda$ is determined by large $\kappa$ values, and therefore by rare events that are less likely to be Gaussian-distributed.
In Figure~\ref{fig:cgf_distributions}, we show the actual distributions of the maximum $\lambda$ values for each case in Figure~\ref{fig:chi2_cgf_kappa_maxcuts}. We see that these are reasonably Gaussian up until about $\lambda=65$, which is consistent with the result seen for the $\chi^2$ tests. Overall, we see that the $\chi^2$ test works best when only the negative-$\lambda$ part of the CGF is sampled ($\lambda_{\rm max}=-5$), and as we will see shortly, this part of the CGF alone is sufficient to provide equally strong cosmological constraints to the PDF. 

There are additional reasons to restrict ourselves to the negative-$\lambda$ region of the CGF. The value of the critical point varies with the angular scale and source redshift considered, and since our theoretical model tends to overpredict its value --- typically because the cylindrical collapse dynamics underpredicts the convergence skewness, and thus underpredicts the high-density tail of the PDF which leads to a larger $\lambda_c$ --- it is safer to take a conservative approach and adopt $\lambda_{\rm max}=0$. On a more theoretical level, the approximation of cylindrical collapse as a proxy for the non-linear dynamics should break down for $\lambda>0$ (and even more when approaching $\lambda_c$) in the CGF  as argued for instance by \cite{Valageas}, although we find it still performs well when comparing to simulations (see Fig. \ref{fig:pdf_cgf}). In the next section, we will explain why we do not expect excluding the positive tail of the CGF to lead to a significant loss of cosmological information. 

\begin{figure}
    \centering
    \includegraphics[width=\columnwidth]{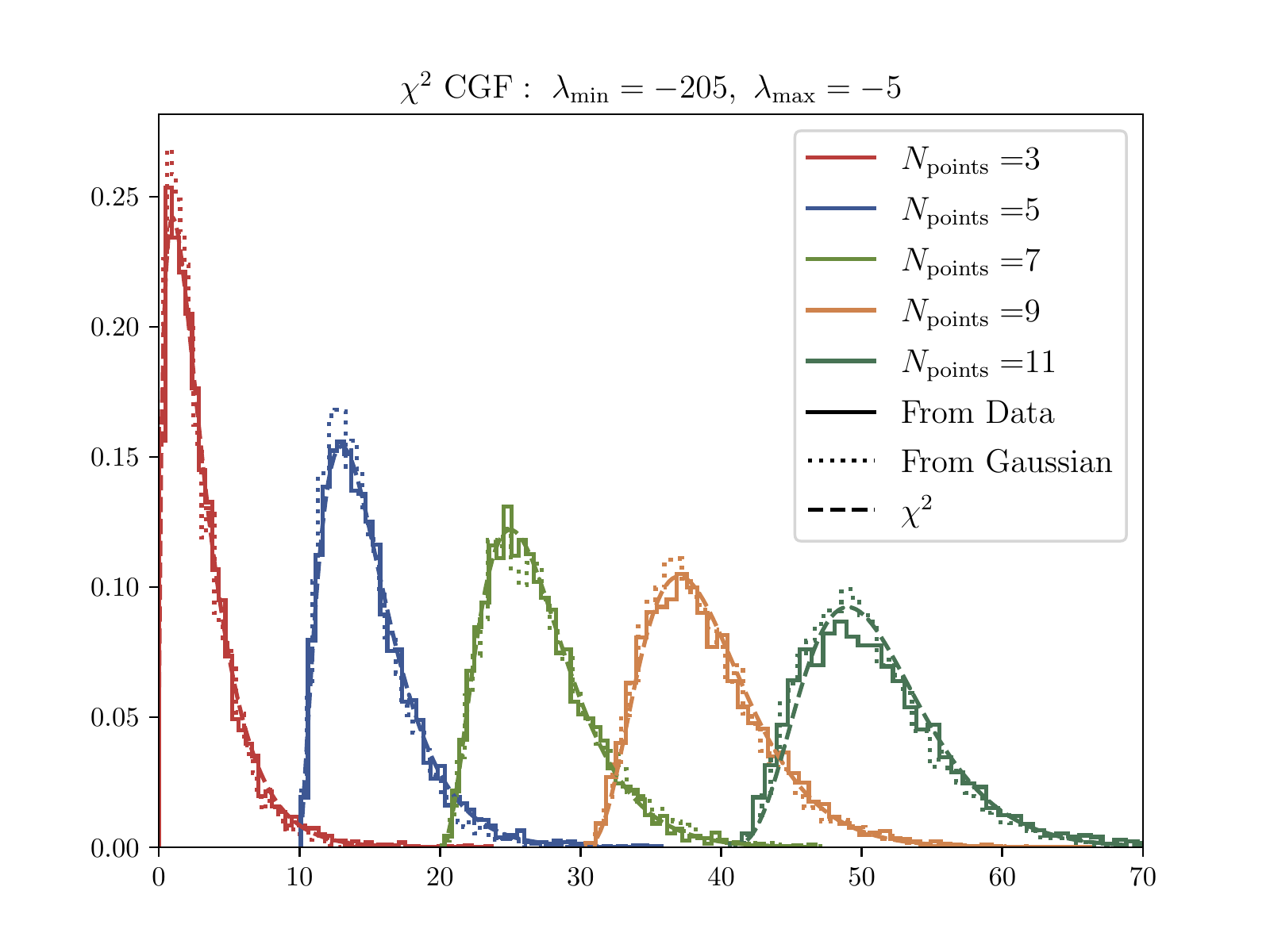}
    \caption{As Figure~\ref{fig:chi2_cgf_kappa_maxcuts}, but in this case the maximum value of $\lambda$ is fixed and the total number of points in the data vector is varied. The $\chi^2$ test seems to become less successful as the number of points is increased. This is a result of the covariance matrix of the CGF being ill-conditioned --- a problem that becomes worse as points are more closely sampled (see main text). This is discussed further in Appendix \ref{app:flask}.}
    \label{fig:chi2_cgf_kappa_maxcuts_equal}
\end{figure}

In Figure~\ref{fig:chi2_cgf_kappa_maxcuts_equal}, therefore, we show $\chi^2$ tests for only the negative part of the CGF ($-205\leq\lambda\leq -5$) while varying the number of $\lambda$ values sampled. The $\chi^2$ works well for a range of samplings, but works less well for 11 points. 
Moreover, we show in Section \ref{sec:fisher} that a Fisher forecast with 11 or more points is unstable. In Appendix \ref{app:flask}, we show that this is due to the covariance matrix being ill-conditioned because of the high level of correlation between points in the CGF. A ridge-regression scheme can be applied to resolve this, as demonstrated in the appendix.

In practice, as we will see in Section~\ref{sec:fisher}, it is nevertheless possible to obtain cosmological constraints competitive with the PDF with quite a small number of data points from the CGF (e.g. three points can provide equivalent constraints to the PDF for two parameters).

\begin{figure}
    \centering
    \includegraphics[width=\columnwidth]{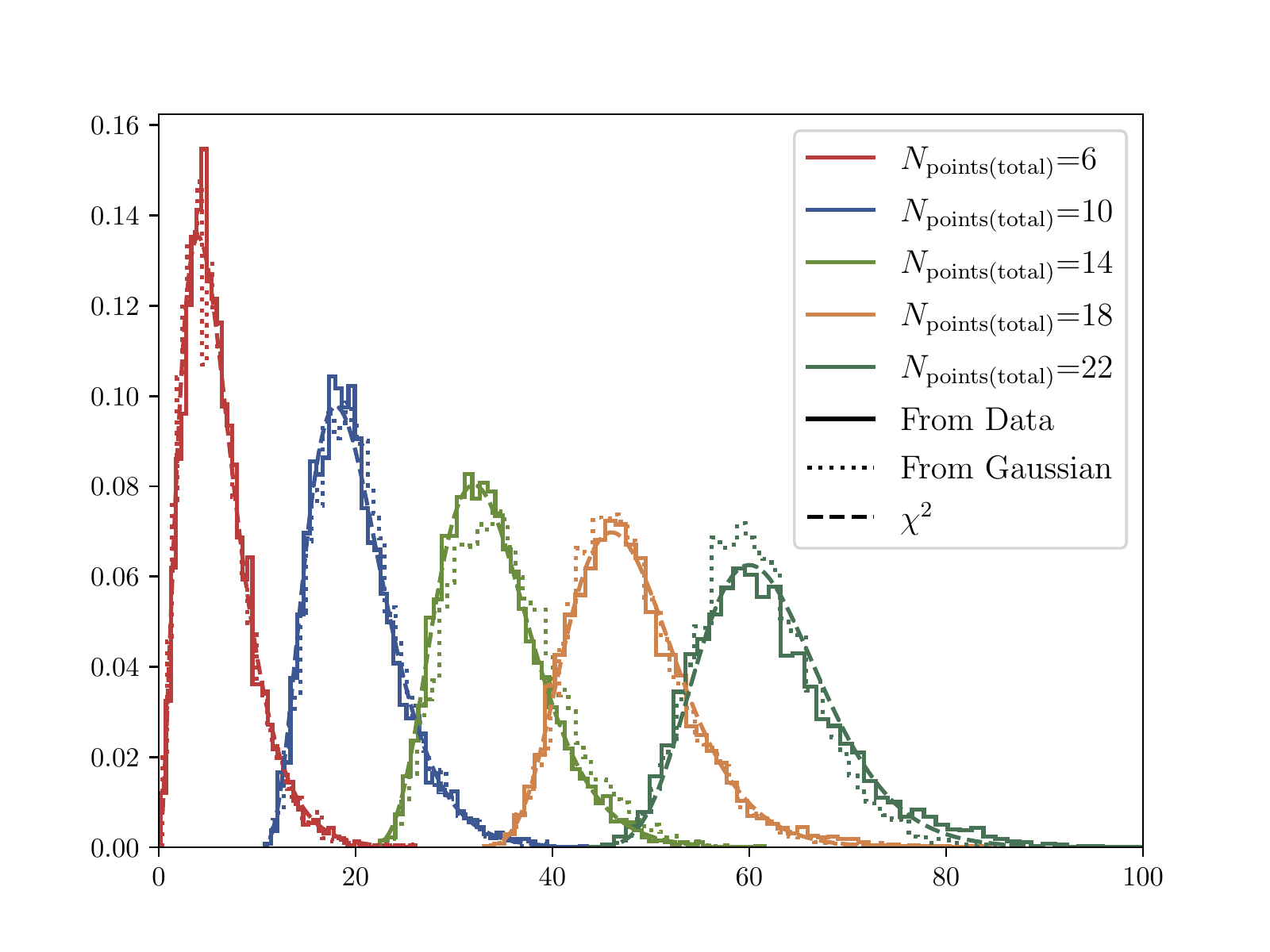}
    \caption{A $\chi^2$ test for a data vector consisting of two CGFs for different smoothing scales ($\theta=7.5^\prime$ and $\theta=10^\prime$). Here $N_{\rm points,total}$ refers to the total number of points sampled from the two CGFs, so $N_{\rm points,total}=6$ refers to three points being sampled from each CGF. We see that the $\chi^2$ test generally performs quite well but somewhat worsens as the number of points is increased.}
    \label{fig:chi2_cgf_2scale}
\end{figure}

Finally, in Figure~\ref{fig:chi2_cgf_2scale}, we show a $\chi^2$ test for a data vector consisting of two CGFs with different smoothing scales. The same values of $\lambda$ are chosen for both CGFs, and the total number of points indicated in the legend refers to the total number of $\lambda$ points sampled over the two CGFs --- for example, $N_{\rm points,total}$ of six means three $\lambda$ points are sampled from each CGF. In this case, we see that the $\chi^2$ test performs well up to and including a total number of sampled points of 14 (seven per CGF).


\subsection{The Negative-$\lambda$ CGF}
\label{subsec::negativeLambda}

In the previous section, we provided a number of reasons for using only the negative tail of the CGF as a cosmological forecasting observable. We can now explain why using only the negative part of the CGF to derive cosmological information should not lead to significant information loss.

\begin{figure}
    \centering
    \includegraphics[width=\columnwidth]{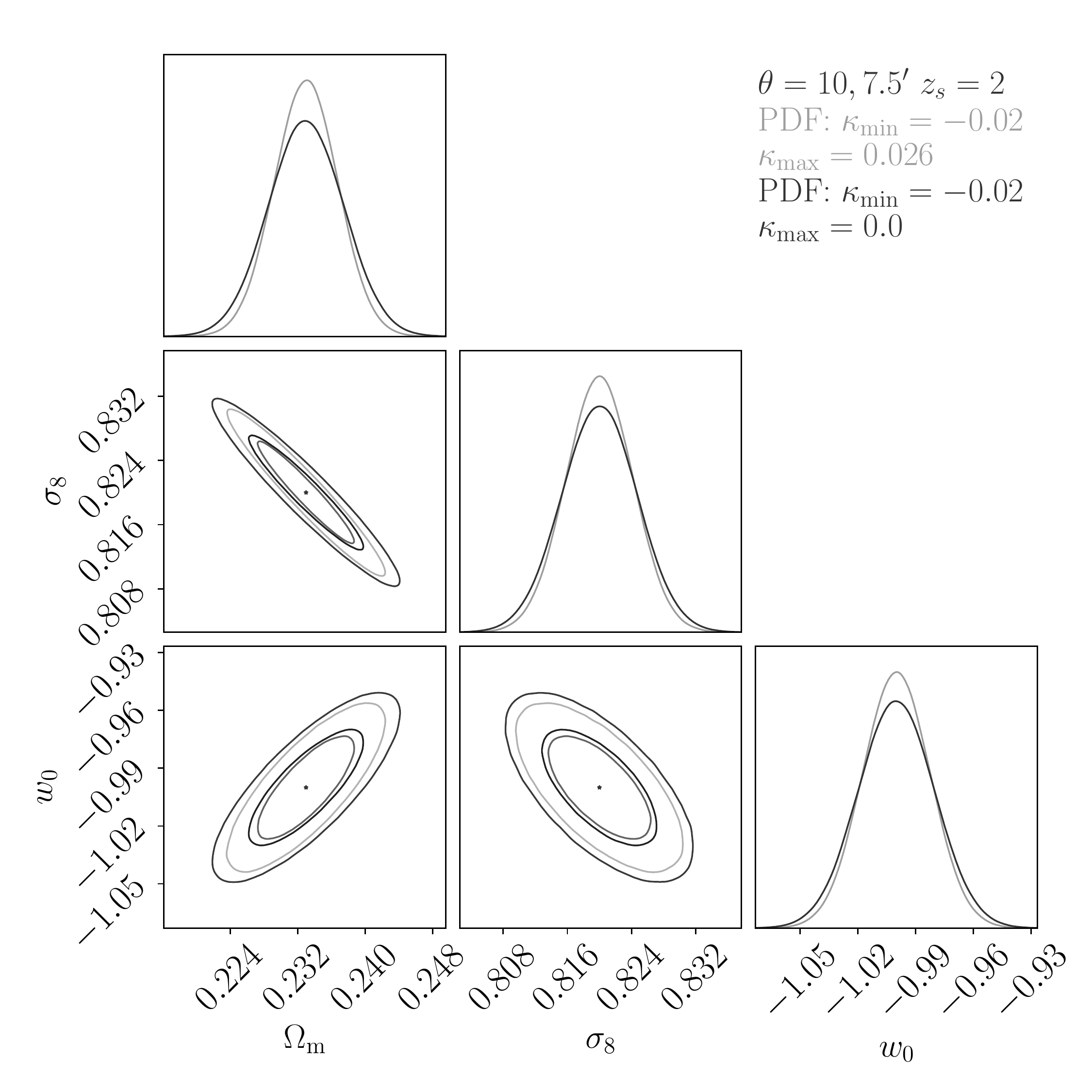}
    \caption{Simple Fisher forecasts for three cosmological parameters (which require combining PDFs from two smoothing scales to provide meaningful constraints) using only half of the PDF (the negative-$\kappa$ part, black) and the full PDF (grey). We see that the constraints from using only half of the PDF are almost equal to those from the full PDF.}
    \label{fig:fisher_pdf_half}
\end{figure}

The first thing to note is that formally, the entire statistical information is located in the vicinity of $\lambda=0$ (one way to understand this is that the cumulants can be defined as the derivatives of the CGF around the zero point), and we can probe this region using the negative tail of the CGF alone. Secondly, according to the $\kappa-\lambda$ relationship shown in Figure~\ref{fig:kappa_lambda}, a cosmological analysis using only the negative-$\lambda$ part of the CGF should be similar (not identical because Figure~\ref{fig:kappa_lambda} is only an approximation) to an analysis only using the negative-$\kappa$ part of the PDF. We show in Figure~\ref{fig:fisher_pdf_half} that cosmological constraints using only the negative-$\kappa$ half of the PDF provide almost equivalent constraints to the full bulk of the PDF. We find that the 1-$\sigma$ constraints on the parameters are only about 10\% weaker when using only the negative half of the PDF. 
We assume that this is because for a positively-skewed PDF, the PDF maximum is located within the negative-$\kappa$ region, so the variance and skewness (the moments from which almost all of the information comes) can be quite well constrained from this portion. We can therefore expect that the constraints from the negative-$\lambda$ CGF capture most of the information one could extract from the $\kappa$-PDF as computed in the LDT framework. Note that this remains a statement about $\kappa$ values in the bulk of the PDF and thus not applicable to cosmological information that could be extracted from extremely non-linear and rare events. 
Moreover, we see in Figure~\ref{fig:pdf_cgf} that LDT seems to underpredict the positive part of the CGF and that the prediction for the negative portion is much more accurate. The same behaviour can be observed at the level of the PDF. This might contribute to the lower weight of the positive part of the CGF.

\subsection{Derivatives with respect to cosmology}

\begin{figure}
    \centering
    \includegraphics[width=\columnwidth]{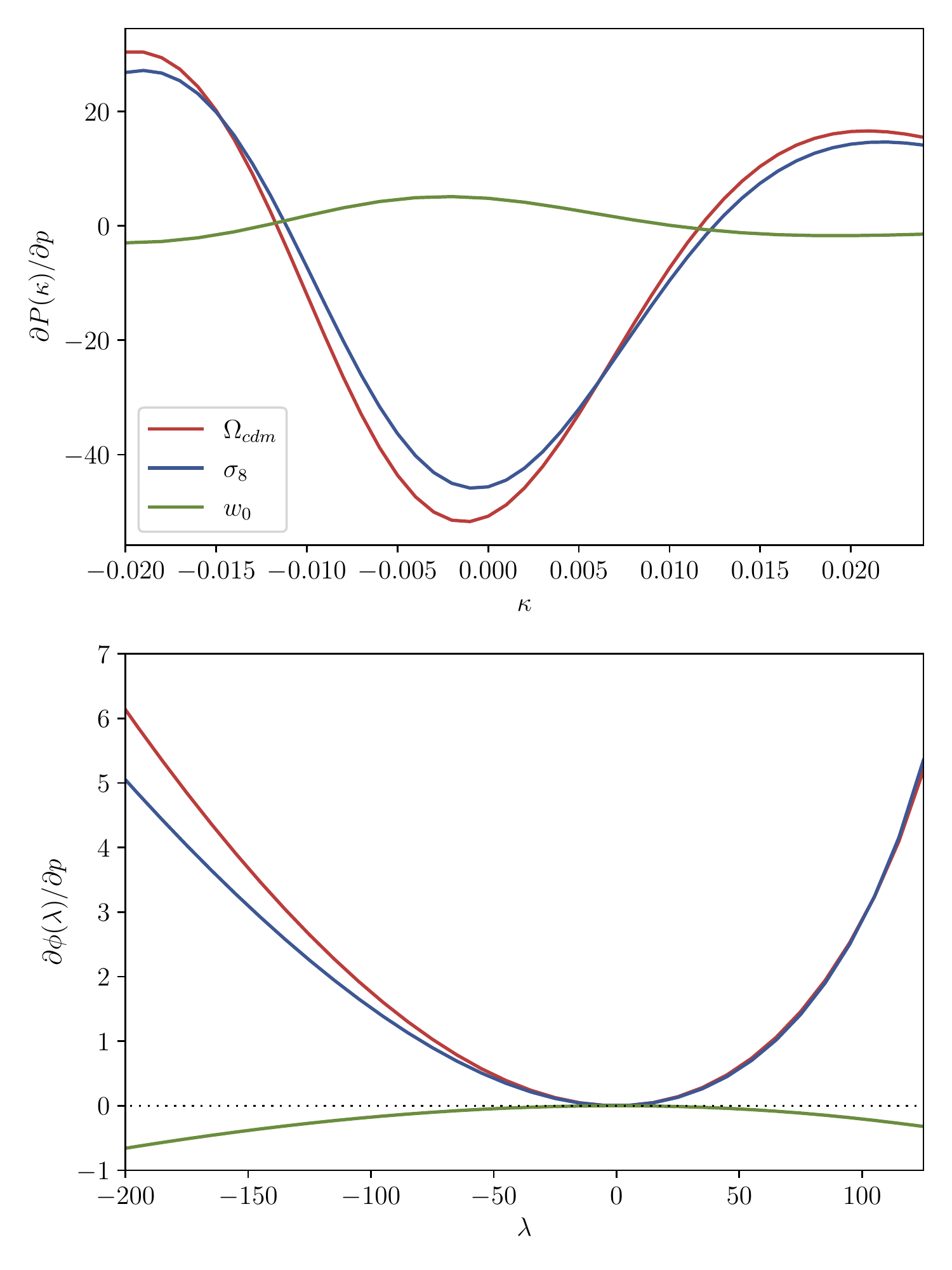}
    \caption{The derivatives of the CGF (upper panel) and PDF (lower panel) with respect to $\Omega_{cdm}$ (red), $\sigma_8$ (blue) and $w_0$ (green) as predicted from the LDT formalism. In both cases, we see that the shapes of the derivatives for $\Omega_{cdm}$ and $\sigma_8$ are very similar, which is consistent with the well-known degeneracy between the two parameters. Note that these derivatives give no indication of the measurability of the relative effects of the cosmological parameters on the observables, and are intended only to show that the relative trends for the PDF and CGF derivatives are consistent.}
    \label{fig:derivatives}
\end{figure}

Figure~\ref{fig:derivatives} shows the derivatives of both the theoretical CGF (lower panel) and PDF (upper panel) with respect to $\Omega_{\rm cdm}$ (red), $\sigma_8$ (blue) and $w_0$ (green). In both cases, we see that the shapes of the derivatives are quite similar for $\Omega_{\rm cdm}$ and $\sigma_8$ --- this reflects the well-known weak lensing degeneracy between these two parameters. We also see in both cases that the primary change is a change in the variance, from the parabola-like derivative for the PDF (an increase in the variance leads to a lower peak and higher tails), and with the derivatives for the CGF looking something like quadratic functions of $\lambda$ in the lower panel.

\subsection{Fisher Forecasts}\label{sec:fisher}

In this section, we perform a number of Fisher forecasts to demonstrate conclusively that the weak lensing convergence CGF can be used to extract an identical amount of cosmological information to the PDF. A Fisher forecast for a set of cosmological parameters, $\bm{\theta}$, given an observed data vector, $\bm{s}$, is defined as
\begin{equation}
F_{ij}= \sum_{\alpha,\beta}\frac{\partial s_\alpha}{\partial \theta_i}C^{-1}_{\alpha \beta}\frac{\partial s_\beta}{\partial \theta_j}~,
\label{eq:Fisher}
\end{equation}
where $s_\alpha$ is element $\alpha$ of $\bm{s}$, the cosmological parameters are indexed as $i,j$ and $C_{\alpha,\beta}$ is the covariance matrix of $\bm{s}$, which is assumed to be cosmology-independent.

The Fisher matrix allows us to determine the error contours on a set of cosmological parameters under the assumption that the likelihood is Gaussian, which should be assured by the relevant $\chi^2$ test. The inverse of the Fisher matrix gives the parameter covariance. The error on the parameter $\theta_i$, marginalised over all other parameters, is given by
\begin{equation}
\delta\theta_i\geq \sqrt{\left(F^{-1}\right)_{ii}} \ .
\end{equation}

As we estimate the covariance matrix in this work from a limited number of covariance maps (108), it is critical to correct for the effect of noise on the matrix inversion process. This is done by multiplying by the Kaufman--Hartlap factor \citep{Kaufman67,Hartlap06},
\begin{equation}
\label{eq:hartlap}
    h=(N_{\rm map} - 2 - N_{\rm points})/(N_{\rm map} - 1).
\end{equation}

\begin{figure}
    \centering
    \includegraphics[width=\columnwidth]{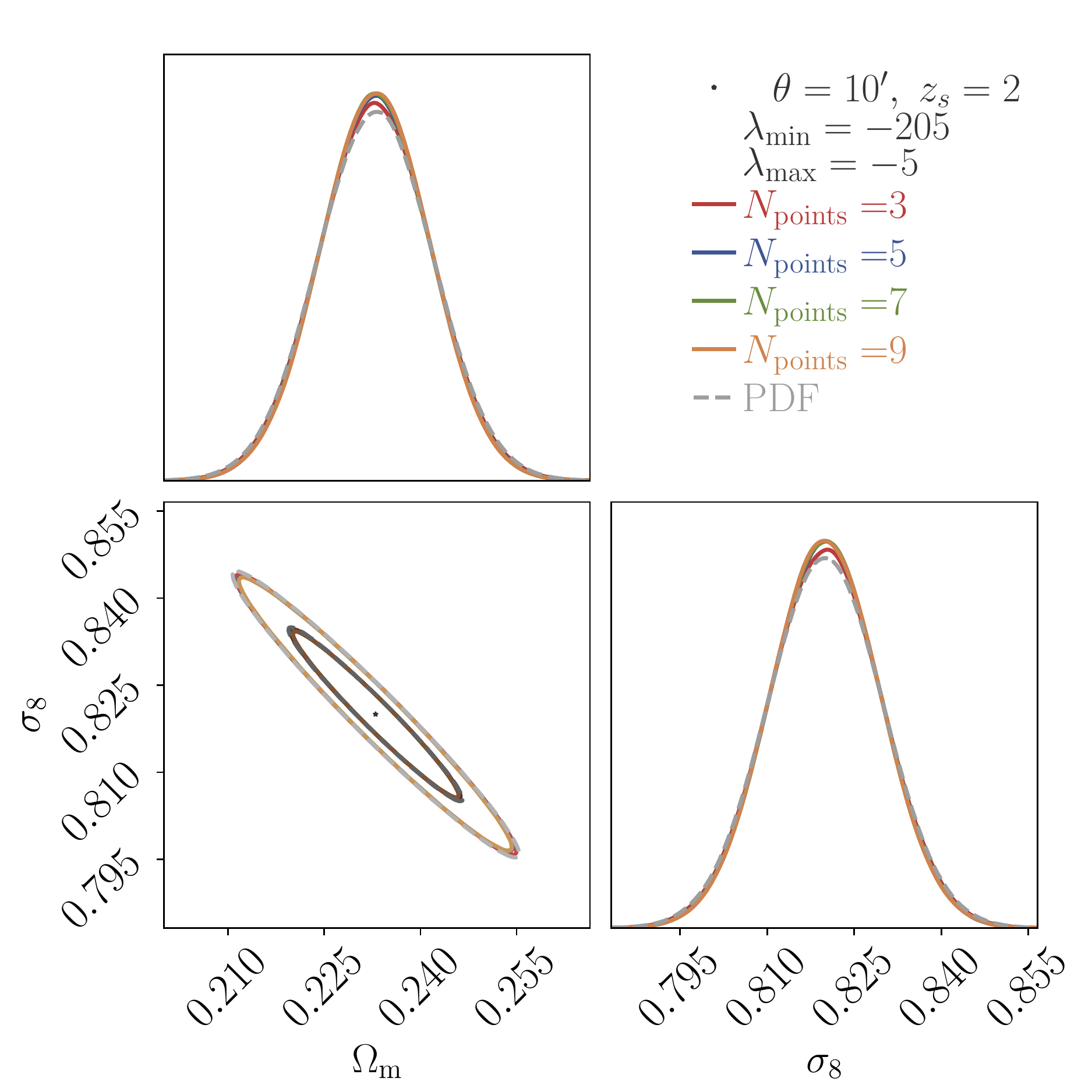}
    \caption{Fisher matrix forecasts for the CGF sampled at several equidistant points between $\lambda_{\rm min}=-205$ (determined from $\kappa_{\rm min}$ in the PDF forecast using the saddle-point approximation) and the final point we measure before zero, $\lambda_{\rm max}=-5$. A comparison forecast for the PDF is shown as a grey dashed line. Sampling three points of the CGF (red) is sufficient to accurately replicate the PDF contour in this simple two-parameter case. The result is consistent as more data points are added (up to nine). Beyond this point, the ill-conditioning of the covariance matrix causes the results to become unreliable --- see Appendix B.}
    \label{fig:fisher_pdf_cgf_multi_varied_varN}
\end{figure}

Here $N_{\rm points}$ can be the number of points sampled for the CGF or the number of bins considered for the PDF, i.e. it is the length of the given data vector. Based on the results of the previous sections, we take a conservative approach and use only the negative part of the CGF to produce Fisher matrix forecasts. In Figure~\ref{fig:fisher_pdf_cgf_multi_varied_varN}, we compare a number of two-parameter Fisher forecasts for the CGF with the corresponding PDF forecast, varying the number of points sampled in the CGF. We see clearly that including only negative values of $\lambda$ up to reasonably close to the zero point produces a result that very closely agrees with the constraint from the PDF (the grey dashed line). Moving beyond the zero point (not shown) tightens these constraints, but we know that this information may not be appropriately distributed from the $\chi^2$ tests in the last section. The minimum value of $\lambda$ is determined from the minimum value of $\kappa$ considered in the PDF forecast using the saddle-point relation between $\kappa$ and $\lambda$, but the results are not very sensitive to this precise choice.

We also see in Figure~\ref{fig:fisher_pdf_cgf_multi_varied_varN} that the CGF forecast performs very well even for a data vector with only three points (at least in this simple two parameter case --- of course, with an increasing number of cosmological parameters, more points will be needed to allow meaningful constraints). The results are consistent up until a data vector of length nine, but beyond this point the agreement breaks down as the $\chi^2$ test fails (see Figure~\ref{fig:chi2_cgf_kappa_maxcuts_equal}). 

\begin{figure}
    \centering
    \includegraphics[width=\columnwidth]{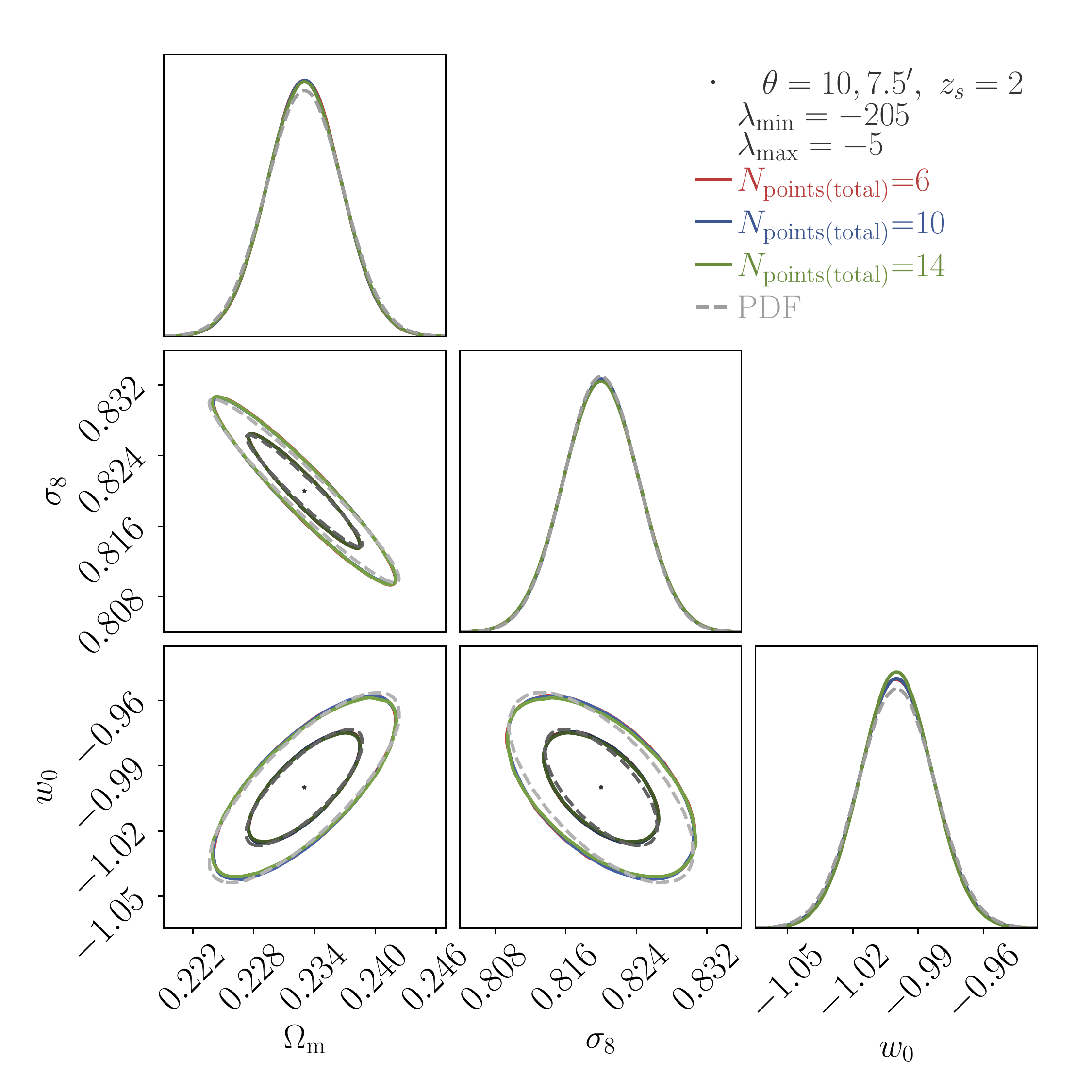}
    \caption{The Fisher matrix forecast expanded to capture three cosmological parameters. In this case, two smoothing scales are combined to provide sufficient information --- 7.5 and 10 arcmin. We see that the results from the CGF and PDF are highly consistent.}
    \label{fig:fisher_pdf_cgf_nparam_2scale}
\end{figure}

In Figure~\ref{fig:fisher_pdf_cgf_nparam_2scale}, we extend to a cosmological forecast for three parameters (adding $w_0$). This requires combining PDFs / CGFs from two different smoothing scales (7.5 and 10 arcmin) to provide meaningful constraints, because the cosmological information from both statistics is provided overwhelmingly by the first two cumulants (effectively the variance and skewness) and two measurements are insufficient to break degeneracies between three parameters. Once again, we see good agreement between the results from the CGF for different samplings and the PDF. 

In general, as already found in \cite{Boyle21}, the cosmological information from convergence one-point statistics is dominated by the information contained in the variance and skewness. However, note that in the case of joint analyses (combining lensing and clustering or different redshift bins), the number of terms required to include all the cumulants of a given order (e.g. order three for the generalisation of the skewness) increases dramatically with the number of random variables. Additionally, cumulants of order higher than three in these scenarios also contain powers of the random variables smaller than three and may therefore contain additional cosmological information. Thus, as advocated throughout this work, it makes more sense to directly use the full CGF in such joint-analyses settings.

\section{Discussion}

\subsection{Applicability of the CGF}

In the previous section, we showed that the CGF can be used as an alternative observable to the PDF that provides roughly equivalent cosmological constraints. In this section, we will discuss in which scenarios the CGF is preferable to the PDF --- in analogy to the complementarity between power spectra and correlation functions in two-point analyses. We recommend considering turning to the CGF in cases where the joint one-point statistics of random variables are considered, either because it can be practically time-consuming and numerically intensive to perform a multi-dimensional inverse Laplace transform to extract the PDF, or physically favourable for the correlation structure between the random variables considered. One example of the former case is the calculation of the joint statistics between different smoothing scales for either matter or lensing fields in the context of LDT. One example of the latter case is considering joint statistics between lensing fields in different tomographic bins or joint statistics between lensing and clustering. In these contexts, the Limber approximation is generally applied, meaning that the slices of the density field along the line of sight are considered statistically independent. This means that random variables are only correlated if they consist of functions of overlapping density slices, which makes their correlation structure extremely straightforward for the CGF (see appendix A of \cite{Barthelemy22}). 

As an example of this complexity, the BNT transform (or nulling strategy) \citep{Nulling} is an invertible linear application that maximally decorrelates lensing fields measured in different tomographic bins. We can consider sets of three random variables correlated two by two, with the first and third variables only being correlated through the second one. This simplified correlation structure is then particularly easily described by relying on CGFs instead of PDFs. This simplified correlation structure also inevitably leads to a simplification in the computation of the full joint statistics, independently of how they are calculated, making it almost mandatory for the efficient estimation of cosmological parameters.

\subsection{Observational and systematic effects}

The observational and systematical issues facing measurements of the CGF are generally the same as for the PDF. Rare events can be excluded from the CGF by choosing a limited range of $\lambda$ values, as for the PDF, in which we generally make cuts in the high and low $\kappa$ tails. The cuts in $\lambda$ chosen can be roughly related to $\kappa$ cuts in the corresponding PDF using the saddle-point approximation, as explained in Section~\ref{kappa-lambda-relation}, just as a rough link can be made between an angular scale cut in a correlation function and a cut in harmonic space $\ell$ in the corresponding power spectrum. As for the PDF, we have seen that the $\chi^2$ test helps one sample the CGF in such a way as to ensure a Gaussian likelihood, and cutting the PDF and CGF based on the $\chi^2$ test leads to consistent cosmological constraints. The addition of statistically-independent statistics like shape noise to the CGF is straightforward --- for example, shape noise can be directly added to the cumulants. Masks in the field of view present only a limited complication to counts-in-cells statistics of direct observables such as the shear field, as shown by \cite{FriedrichDES17}, because masks do not meaningfully affect local quantity measurements. Masks do, however, complicate the reconstruction of the convergence field, but it has been at least empirically shown on numerical simulations that the effect of this on the scaled cumulant generating function (equation~\eqref{defscgf}) is minimal \citep{GattiDESsim,gattiDES}, such that mask effects can be captured through their impact on the variance.
This suggests a possible advantage to making use of the CGF as the basis for analysis of the convergence field.

Finally, in the hypothetical absence of a theoretical model, one possible disadvantage of the CGF compared to the PDF would be that the CGF likelihood tends to be less Gaussian than that of the PDF in small area maps (e.g. $\sim$ 25 square degrees). This means a large set of large-area simulated maps would be required to calculate the covariance. In addition, the fact that the elements of the CGF covariance matrix are very correlated can make it more prone to numerical issues in its estimation and inversion, as discussed in Appendix \ref{app:flask}. However, note that we also demonstrate in Appendix \ref{app:flask} that in practice, fast lognormal models can also be used to estimate the covariance matrix of the CGF.

\section{Conclusions}

The value of one-point statistics as a complement to two-point statistics has been clearly documented \citep[e.g.][]{FriedrichDES17, Uhlemann_2020, Boyle21}. Besides carrying additional cosmological information, one-point statistics have a number of valuable properties: they are easily measured and can be theoretically modelled from first principles, it is straightforward to remove the influence of very non-linear or rare events, and it is easy to only consider the highest signal-to-noise part of the signal. Up until now, works aiming to exploit the full hierarchy of one-point cumulants of cosmological fields have focused on the probability distribution function (PDF).

In this work, we have used the weak lensing convergence as an example to demonstrate for the first time how the CGF can be used as an alternative cosmological observable as a probe of one-point statistics. We have shown that the likelihood of the CGF restricted to the negative axis is Gaussian and that a sampling of this section of the CGF can be used to extract the same amount of cosmological information as is contained in the bulk of the corresponding PDF. We have performed simple Fisher matrix forecasts to demonstrate this agreement, leveraging theoretical predictions for how the CGF depends on cosmology. 

The key figures that summarise our results are as follows. In Figure~\ref{CGF_from_cut_PDF}, we showed that the range of the CGF we sample for our Fisher matrix forecasts is very insensitive to rare events. In Figure~\ref{fig:chi2_cgf_kappa_maxcuts_equal}, we showed that the data vector we consider is Gaussian distributed and therefore suitable for Fisher matrix forecasts. In Figures~\ref{fig:fisher_pdf_cgf_multi_varied_varN} and \ref{fig:fisher_pdf_cgf_nparam_2scale}, we showed that equivalent results can be obtained from Fisher forecasts using either the PDF or CGF, including for multiple parameters and combinations of smoothing scales.

The aim of this work has been to provide proof of concept for the exploitation of the CGF as a cosmological probe. The most logical next step would be to approach more complex scenarios in which we expect the CGF to be particularly useful, including tomographic analyses and studies of cross-correlated lensing and galaxy clustering fields. Extending our formalism to the galaxy density field would require a treatment of galaxy bias and shot-noise. At the level of the 1-point PDF these have been incorporated into the LDT framework by \cite{friedrich_2021} and their results can be translated to the CGF level. Another natural extension would be the application of our theoretical model of the CGF to real data --- the efficiency with which the CGF can be modelled using LDT makes it particularly appealing for MCMC analyses. In this case, one would need to ensure the survey-specific accuracy of the theoretical model to avoid a survey-dependent bias in the estimation of cosmological parameters and their confidence intervals. We emphasise that the work presented here is not intended to represent a realistic CGF-based analysis that can be applied to survey data. A primary purpose for validating the feasibility of CGF-based analyses is to allow for more efficient evaluations of a `nulled' 1-point statistic, the result of applying the BNT transform \citep{Nulling} to a combination of tomographic measurements with appropriate weights to cancel out small-scale contributions. Nulling has been shown to dramatically improve the accuracy of the theoretical PDF from LDT \cite{Barthelemy20a} and is the foundation for using CGFs as a way to reduce the complexity of the numerical computation of the theory data vector in a tomographic setting \citep{Barthelemy22}. In practical applications, we envision the use of the CGF as a probe of cosmological parameters  in conjunction with 2-point statistics and 3$\times$2 point analyses.

\section*{Acknowledgements}

ABo and SC's work is supported by the SPHERES grant ANR-18-CE31-0009 of the French {\sl Agence Nationale de la Recherche}, ABa's by the ORIGINS excellence cluster and SC's by Fondation MERAC. CU is supported by the STFC Astronomy Theory Consolidated Grant ST/W001020/1 from UK Research \& Innovation. This work has made use of the Infinity Cluster hosted by Institut d'Astrophysique de Paris. We thank St\'ephane Rouberol for running smoothly this cluster for us. We thank Francis Bernardeau for helpful discussions and comments. ABo thanks L. Boyle and O. Boyle for fruitful discussions. 

\section*{Data Availability}

There is no new data associated with this article.

\bibliographystyle{mnras}
\bibliography{references} 

\appendix

\section{Validity and bias of the CGF estimator}\label{app:estimator}

The literature from fields making use of the cumulant generating function \citep[see for example][]{math1,math2} does not to our knowledge make use of a more refined estimator for the cumulant generating function than the one we use throughout this paper.\footnote{There exist, however, unbiased estimators for the individual cumulants called \href{https://mathworld.wolfram.com/k-Statistic.html}{k-statistics} that may be used to correct the CGF to leading order. We believe the following shows this is not necessary for current and next generation galaxy surveys.}  However, though the estimator for the moment generating function $\langle e^{\lambda \kappa}\rangle$ is unbiased, taking the natural logarithm of this estimator does not lead to an unbiased estimator of the log of the moment generating function (the CGF). 
The goal of this section is thus to more quantitatively estimate the (leading order) bias of the CGF estimator and show that it is negligible for our purposes. We denote $\hat{\phi}$ and $\hat{M}$ our estimators for respectively the CGF and MGF. To leading order, the CGF bias can be directly related to the variance of $\phi(\lambda)$ as:
\begin{align}
    \hat \phi &= \ln (\hat M)\\
    &\approx \ln (M +\hat{\delta M})\\
    &\approx \ln M + \ln \left(1 +\frac{\hat{\delta M}}{M}\right)\\
    &\approx \phi + \frac{\hat{\delta M}}{M} - \frac{1}{2}\left(\frac{\hat{\delta M}}{M}\right)^2.
\end{align}
And since the estimator for the MGF is unbiased, $\langle \hat{\delta M} \rangle = 0$, and
\begin{align}
    \Rightarrow \langle \hat \phi - \phi  \rangle &= -\frac{1}{2}\left\langle \left(\frac{\hat{\delta M}}{M}\right)^2 \right\rangle\\
    &= -\frac{1}{2}\left\langle \left(\hat{\delta \phi}\right)^2 \right\rangle\\
    &= -\frac{1}{2}\mathrm{Var}(\hat\phi).
    \label{bias_ana}
\end{align}
The bias thus highly depends on the survey size\footnote{Note especially that the bias relative to the standard deviation $\frac{\langle \hat \phi - \phi  \rangle}{\mathrm{Std}(\hat\phi)} \approx -\frac{1}{2}\mathrm{Std}(\hat\phi) \propto \sqrt{\frac{1}{A_{\mathrm{survey}}}}$.} but is easily given by the diagonal of the covariance matrix.

To confirm our derivation, we also measure from the Takahashi simulation the jackknife (leading order) bias of the CGF estimator. We thus measure the unbiased estimator of the MGF from the $n = $108 Takahashi convergence maps used throughout this work, smoothed on a scale of $10$ arcmin. Our estimator of the CGF is then given by
\begin{equation}
    \hat{\phi}(\lambda) = \log\left(\frac{1}{n}\sum_{i=1}^{i = n} \hat{M}_i(\lambda)\right).
\end{equation}

We also define $n$ different estimators for the CGF, each given by
\begin{equation}
    \hat{\phi}_j(\lambda) = \log\left(\frac{1}{(n-1)}\sum_{i=1; i\neq j}^{i = n} \hat{M}_i(\lambda)\right),
\end{equation}
and the \textit{jackknife} estimator, which is given by
\begin{equation}
        \hat{\phi}_{\rm jack}(\lambda) = \frac{1}{n}\sum_{j=1}^{j = n} \hat{\phi}_j(\lambda).
\end{equation}

From the jackknife estimator, we can obtain the exact leading order bias ($\mathcal{O} \frac{1}{n}$), which is given by
\begin{equation}
    {\rm bias}_{\phi} = \langle\hat{\phi} - \phi\rangle = n \left((n-1) (\hat{\phi}_{\rm jack}-\hat{\phi})\right),
    \label{CGFbias}
\end{equation}
where the overall $n$ factor comes from the fact that the bias is estimated for a survey size of $n$ independent full-skies and we want the bias for only one.

\begin{figure}
    \centering
    \includegraphics[width=\columnwidth]{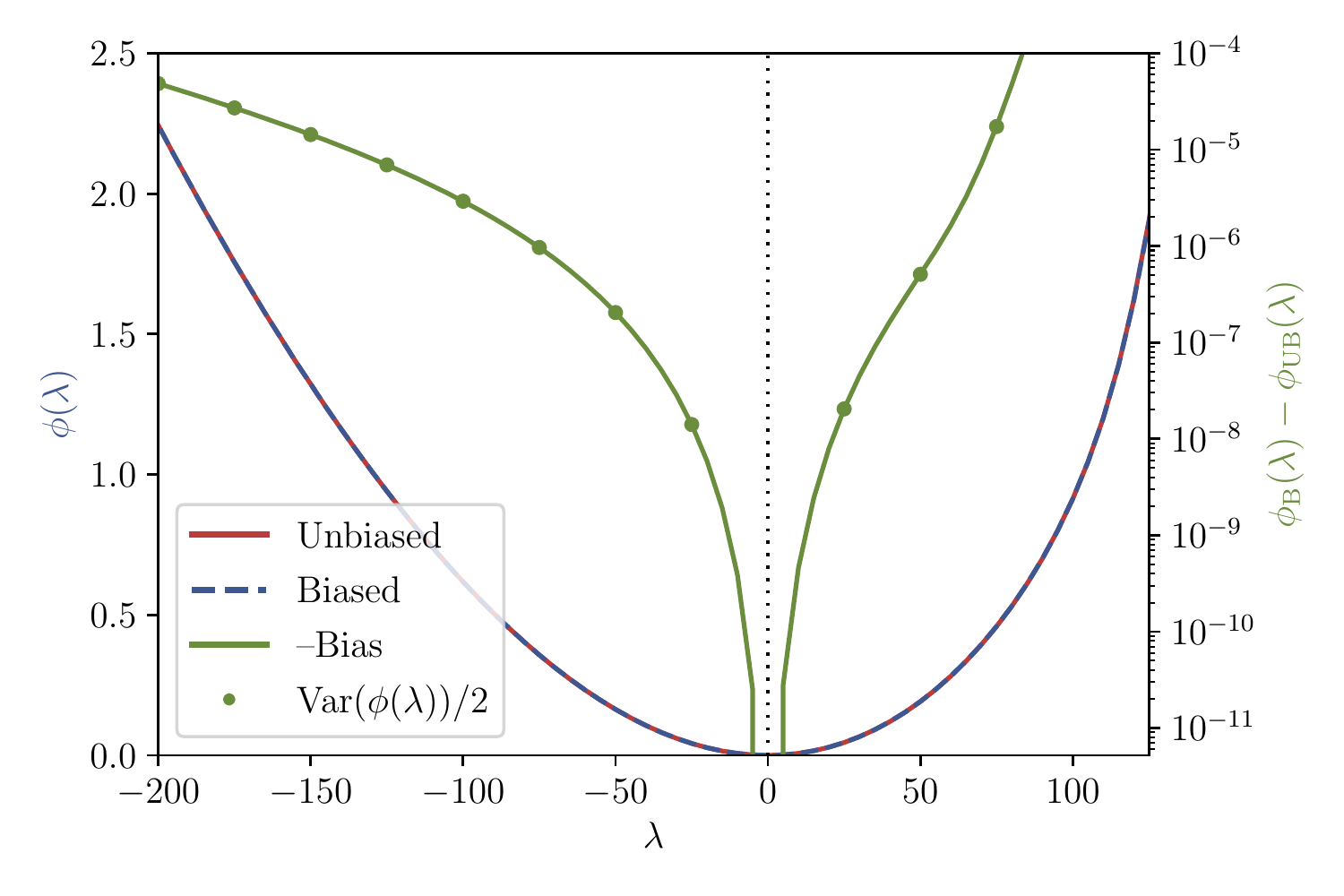}
    \caption{The unbiased (red) and biased (blue) CGF estimators (see the left y-axis). The bias itself is shown in green 
 (see the right y-axis). We see that the bias is totally negligible in the range of $\lambda$ values we consider in our analysis. The green points show half the variance of $\phi(\lambda)$, i.e. the diagonal in Figure~\ref{fig:covariance} divided by a factor of two. As demonstrated in the text, this matches the bias very well.}
    \label{fig::CGFbias}
\end{figure}

Thus using either equation~\eqref{bias_ana} or \eqref{CGFbias}, we can correct our estimator of the CGF and determine the magnitude of the effect on the overall measurement. In Figure~\ref{fig::CGFbias}, we first compare (green line and points) the two expressions we obtained for the bias and see that they are virtually identical. We also plot the two resulting estimators of the CGF --- biased and unbiased. We see that the bias for a full-sky study is totally negligible, and could easily be corrected for smaller area surveys.

\section{COVARIANCE MATRIX CONDITIONING}\label{app:flask}

\begin{figure}
\includegraphics[width=\columnwidth]{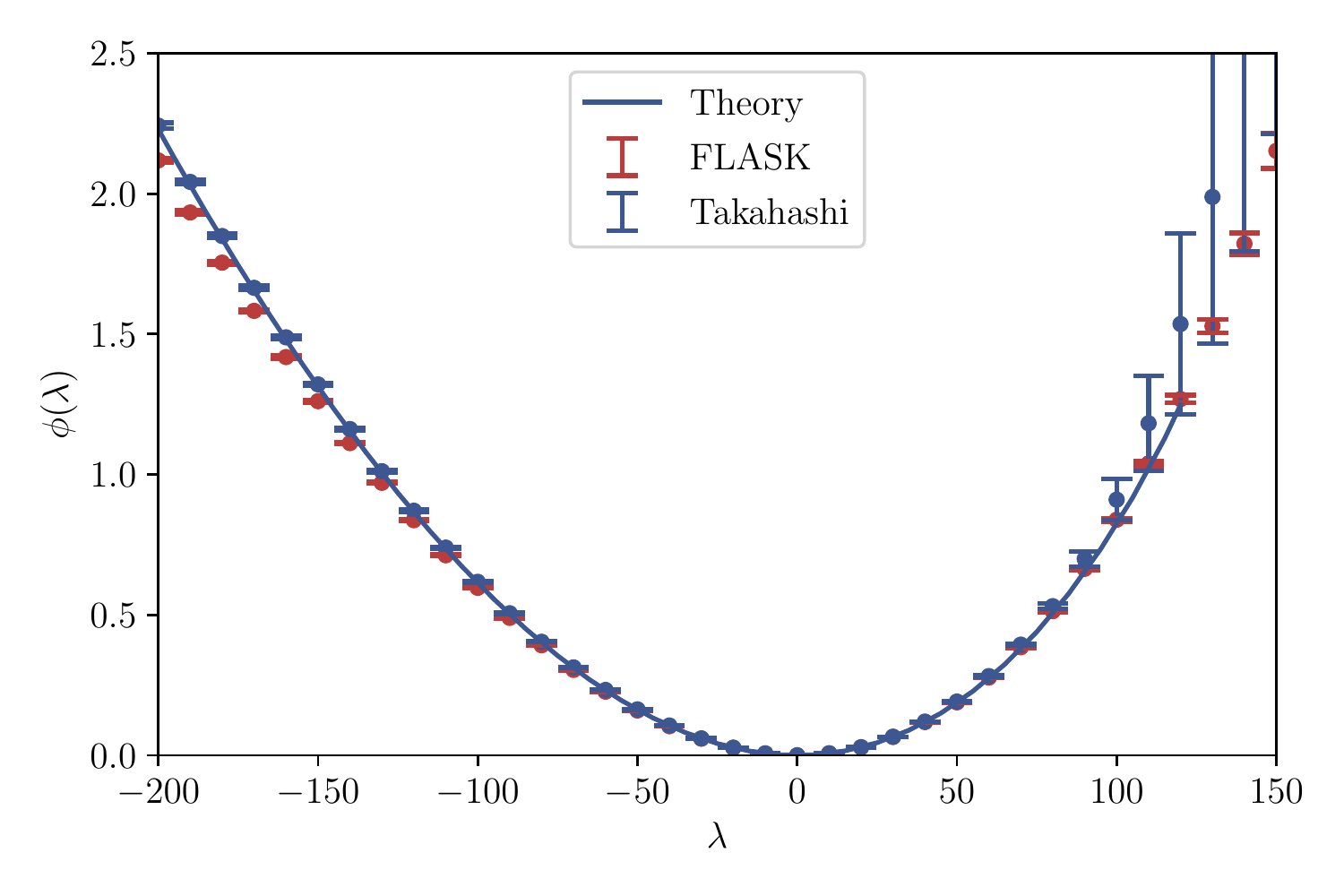}
\caption{A comparison between the mean CGF from FLASK (from 100 shifted lognormal convergence maps) and the theoretical CGF for the same cosmology from LDT.}\label{fig:fiducial_cgf_flask}
\end{figure}

\begin{figure}
\includegraphics[width=\columnwidth]{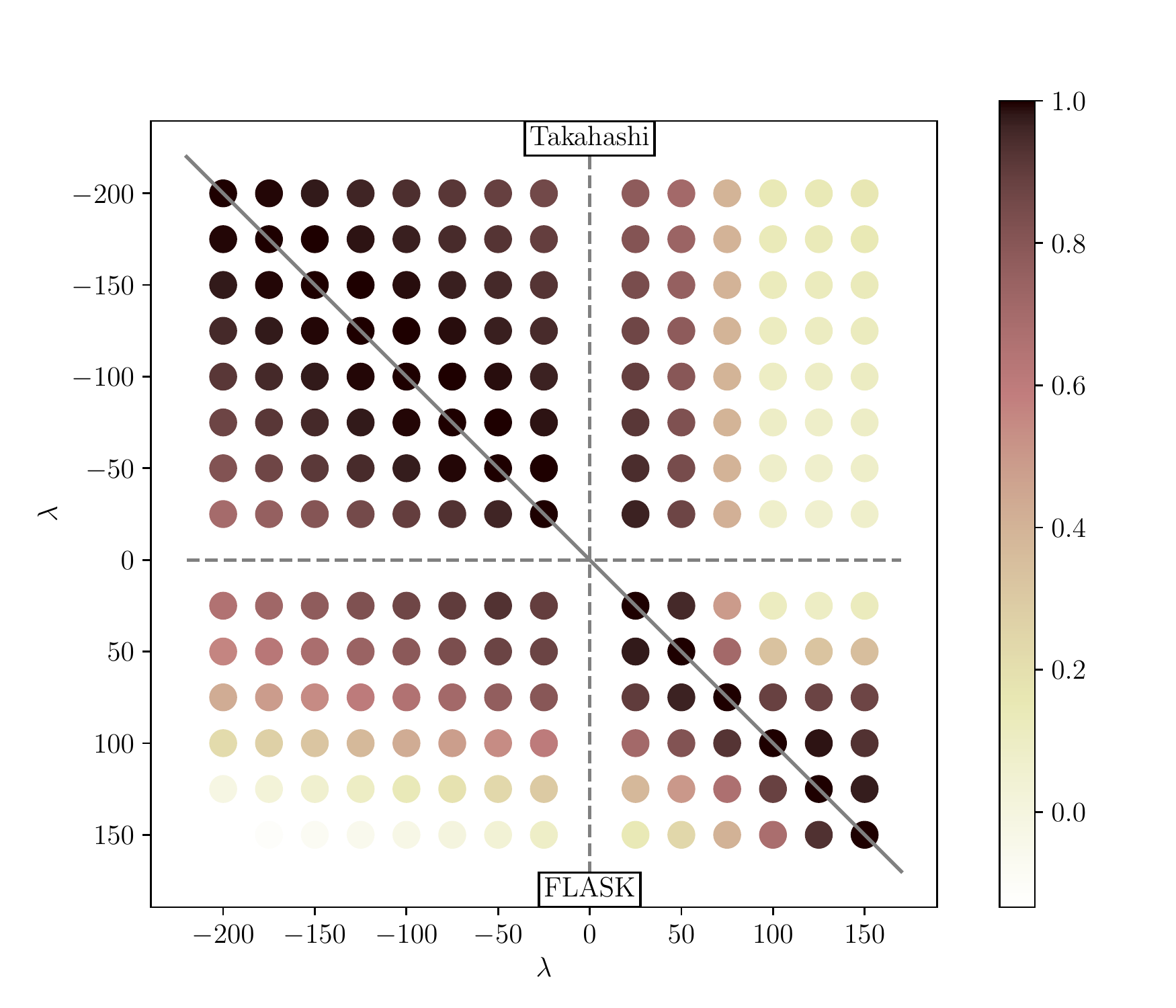}
\includegraphics[width=\columnwidth]{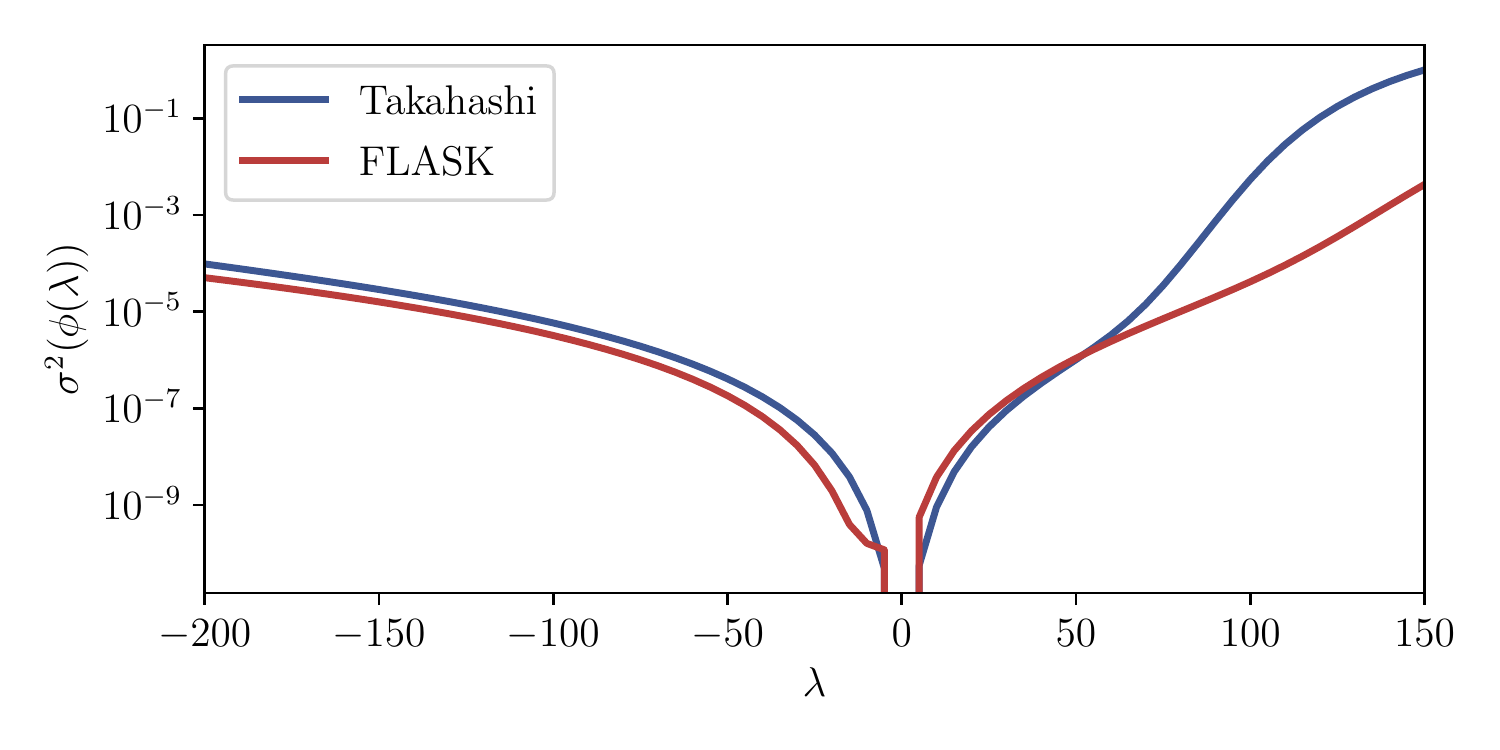}
\caption{(Upper panel) The correlation matrices of the CGF from the Takahashi simulations (top right) and the FLASK code (bottom left). Both are extracted from 108 maps. (Lower panel) The diagonals of the corresponding covariance matrices. There are clear differences between the two covariance matrices for $\lambda>0$ (including anti-correlations between points in the FLASK case), but the agreement for $\lambda<0$ is reasonably good.}\label{fig:fiducial_covariance_lambdas_flask}
\end{figure}

In previous work \citep{Boyle21}, we made use of the FLASK code \citep{Flask} to rapidly generate large numbers of lognormal convergence maps to rapidly produce covariance matrices for the PDF. The CGF is undefined for positive $\lambda$ in a strictly lognormal field, but the smoothing we impose on the FLASK maps, along with the original finite resolution and sky area, cause a deviation from a perfect lognormal distribution. In any case, we have conservatively used only the negative-$\lambda$ CGF throughout most of this paper.

In Figure~\ref{fig:fiducial_cgf_flask}, we compare the mean CGF from 100 FLASK realisations to the theoretical prediction from LDT. We see reasonably good agreement, especially considering that the FLASK realisations underpredict the kurtosis by construction. 
The lognormal fields generated by FLASK are defined by the input angular power spectrum \citep[which we generate using the Boltzmann code CLASS,][]{CLASS} and a shift parameter that is a function of the desired skewness (measured directly from the smoothed Takahashi simulation maps). The shift parameter is given by

\begin{align}\label{eq:xavier_lambda}
    \lambda &= \frac{\sigma}{\tilde{\mu}_3}\left(1+y(\tilde{\mu}_3)^{-1}+y(\tilde{\mu}_3)\right)-\langle \kappa\rangle, \\
    y(\tilde{\mu}_3) &= \sqrt[3]{\frac{2+\tilde{\mu}_3^2+\tilde{\mu}_3\sqrt{4+\tilde{\mu}_3^2}}{2}}\,, \notag
\end{align}
where $\langle \kappa\rangle$ is the mean (zero for the $\kappa$ field), $\sigma$ is the variance and $\tilde \mu_3$ is the skewness. We calculate the shift parameter for the Takahashi cosmology at $z_s=2$ to be 0.085.

In Figure~\ref{fig:fiducial_covariance_lambdas_flask}, we show correlation matrices generated from the sets of Takahashi and FLASK convergence maps (108 maps each). We see that the agreement in the negative-$\lambda$ region seems reasonably strong by eye, but there are clear differences for positive $\lambda$. We also see anti-correlations between negative and positive $\lambda$ values in the FLASK case, while all of the points in the Takahashi case are positively correlated. The lower panel of Figure~\ref{fig:fiducial_covariance_lambdas_flask} shows the diagonals of the corresponding covariance matrices. We see reasonable agreement for negative $\lambda$ (taking into account that there are some differences in the mean CGFs as seen in Figure~\ref{fig:fiducial_cgf_flask}). 

In total, we generated 500 FLASK convergence maps with our specifications. One might expect that a covariance matrix generated from this significantly larger number of maps would allow for accurate Fisher matrix forecasts with a finer sampling of the CGF than was possible with the original 108 Takahashi maps. However, we find that the covariance matrices for the CGF are generally ill-conditioned, with very large condition numbers once the sampling of the CGF reaches a certain density. This is not an issue for the PDF but is somewhat predictable for the CGF, because the points in the CGF are very strongly correlated with each other. The condition number of the CGF covariance matrices is found to decrease only minimally even when significantly increasing the number of maps used. Naturally, the magnitude of this problem increases considerably as one considers more and more points from the CGF, as points that are closer together in $\lambda$ are even more strongly correlated.

There are a number of approaches one can take to improve the conditioning of a covariance matrix, such as restructuring the data vector into a form that decreases the correlation between its individual components. However, the simplest approach seems to be to define a small positive constant $\alpha$ multiplied by the identity matrix that we add to the estimated covariance matrix.  This has the effect of ensuring the eigenvalues are sufficiently different from zero for a stable numerical inversion. This process, called \lq ridge regression\rq, has been exploited in a few other works for other cosmological observables, e.g. \cite{ridge_regression}. The constant $\alpha$ should be chosen from the numerical error we expect on the small eigenvalues of our covariance matrix so that the shift we impose just ensures they are no longer compatible with zero. To estimate it, we compute the standard deviation $\hat{\sigma}_{\rm jack}$ of the jacknife estimator $\hat{J}_{\rm ack}$ of the eigenvalues of the covariance matrix and determine the minimal $\alpha$ value for which $\hat{J}_{\rm ack} - \hat{\sigma}_{\rm jack}$ is always positive. Using 108 jackknife replicates based on the 108 Takahashi realisations, we find that a good value is $\alpha \sim 5\cdot10^{-16}$. Finally, we note that since this method increases the smallest eigenvalues of the covariance matrix, it can only decrease the final constraints on the cosmological parameters even if the value of $\alpha$ is not ideal. This makes this a conservative approach for cosmological analysis. 

\begin{figure}
    \centering
    \includegraphics[width=\columnwidth]{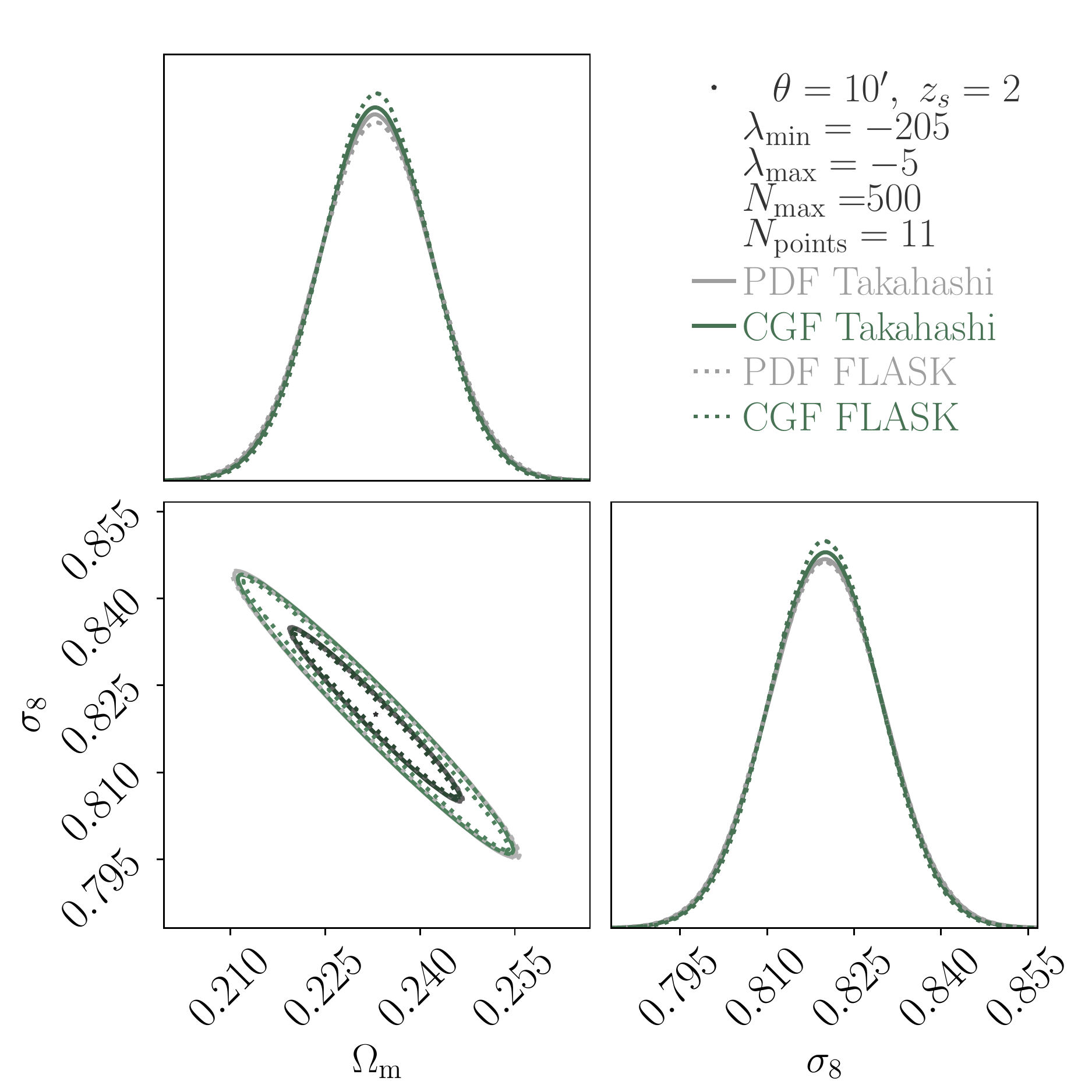}
    \caption{Fisher forecasts for the PDF (grey) and CGF (green) using the covariance matrix constructed from 108 Takahashi convergence maps (solid) and 500 FLASK lognormal convergence maps (dotted). Eleven points are sampled from the CGF between $\lambda_{\rm min}$ of -205 and zero. We take steps to reduce the conditioning number of the CGF covariance, which allows it to be stably inverted (see main text), even for a fine sampling of points of the CGF. We see that the Fisher forecasts are all mutually consistent --- the PDF and CGF provide approximately the same constraints, as do the Takahashi and FLASK covariances.}
    \label{fig:fisher_flask}
\end{figure}

In Fig. \ref{fig:fisher_flask}, we show Fisher forecasts for the PDF (grey) and CGF (green) using both the Takahashi (solid, 108 maps) and FLASK (dotted, 500 maps) covariance matrices. The approach described above is used to improve the conditioning of the CGF covariance matrices, allowing us to produce consistent Fisher forecasts even when sampling a large number of points (11) from the CGF. We see excellent agreement between all four cases. This also confirms that the FLASK code can be used to rapidly generate covariance matrices for the weak lensing convergence CGF, which is helpful because such covariances can be combined with theoretical derivatives (e.g. from LDT) to produce cosmological forecasts without the need for any input from expensive N-body simulations.

\label{lastpage}

\end{document}